\documentclass[twocolumn]{aastex631}

\usepackage{graphicx}	
\usepackage{amsmath}	
\usepackage{xcolor}
\usepackage{graphicx}
\usepackage{caption}

\DeclareCaptionFormat{cont}{#1 (cont.)#2#3\par}
\usepackage[normalem]{ulem}

\def\hb{H$\beta$}

\def\rfeop{R$_{\rm 4570}$}
\def\rfenir{R$_{\rm 1\mu m}$}

\def\feii{Fe\,{\sc ii}}

\def\kms{km s$^{-1}$}
\def\cloudy{{\sc cloudy}}

\newcommand{\deni}[1]{{\color{blue} Denimara:}}
\newenvironment{myfont}{\fontfamily{lmtt}\selectfont}{\par}
\DeclareTextFontCommand{\textmyfont}{\myfont}

\begin{document}

\title{Simultaneous modeling of Fe\,{\sc ii} emission in the optical and near-infrared in a prototypical Narrow-Line Seyfert 1 galaxy}

\author[0000-0003-4153-4829]{Denimara Dias dos Santos}
\affiliation{Laborat\'orio Nacional de Astrof\'isica (LNA), Rua dos Estados Unidos 154, Bairro das Na\c c\~oes, Itajub\'a, MG, Brazil\\}
\affiliation{ Divis\~ao de Astrof\'isica, Instituto Nacional de Pesquisas Espaciais (INPE), Av. dos Astronautas 1758, S\~ao Jos\'e dos Campos, SP, Brazil  \\}

\author[0000-0002-5854-7426]{Swayamtrupta Panda}
\altaffiliation{Gemini Science Fellow}
\affiliation{International Gemini Observatory/NSF NOIRLab, Casilla 603, La Serena, Chile\\}
\affiliation{Laborat\'orio Nacional de Astrof\'isica (LNA), Rua dos Estados Unidos 154, Bairro das Na\c c\~oes, Itajub\'a, MG, Brazil\\}

\author[0000-0002-7608-6109]{Alberto Rodríguez-Ardila}
\affiliation{Observat\'orio Nacional, Rua Jos\'e Cristino, 77, S\~ao Cristov\~ao, 20921-400, Rio de Janeiro, RJ, Brasil}

\author[0000-0001-9719-4523]{Murilo Marinello}
\affiliation{Laborat\'orio Nacional de Astrof\'isica (LNA), Rua dos Estados Unidos 154, Bairro das Na\c c\~oes, Itajub\'a, MG, Brazil\\}
\affiliation{International Gemini Observatory/NSF NOIRLab, Casilla 603, La Serena, Chile\\}



\begin{abstract}
This work investigates the \feii{} emission in active galactic nuclei (AGN), combining observational data from optical and near-infrared (NIR) spectra of the prototypical \feii{} emitter I~Zw~1 with state-of-the-art photoionization modeling. Using updated \feii{} atomic datasets (\citealt{smith19, tayal2018, Bautista_2015}), we explore a wide parameter space 
to determine the physical conditions of \feii{}-emitting regions in the broad-line region (BLR). Our results show that optical (\rfeop{}) and NIR (\rfenir{}) \feii{} emissions can be simultaneously reproduced under consistent conditions, with the best agreement using the \citet{smith19} dataset, for hydrogen densities of $10^{11.0}$ to $10^{12.0}$ cm$^{-3}$ and near-solar metallicity. We quantify, for the first time, the impact of Lyman-$\alpha$ fluorescence through physical conditions for \feii{} emissions in both regimes, revealing its dominant role in the NIR and, in contrast, highlighting a stronger influence of collisional processes in the optical. Additionally, for the first time, we compare optical and NIR \feii{} emissions simultaneously with O\,{\sc i} and Ca\,{\sc ii} triplet (CaT), reinforcing their similar spatial region, physical properties, and their usability as better proxies for optical \feii{}. Our findings support the idea of a vertical BLR structure, with NIR \feii{} and O\,{\sc i} originating in less dense regions in the cloud than optical \feii{} and CaT. 

\end{abstract}



\section{Introduction} \label{sec:intro}

Active Galactic Nuclei (AGNs) are energetic phenomena located at the centers of galaxies, characterized by a supermassive black hole (SMBH) in the process of accreting matter. Surrounding the central source are clouds directly exposed to the AGN ionizing continuum, forming the Broad Line Region \citep[BLR;][]{Wills1985BroadEmission, Joly1993AnLines}. This compact region produces broad emission lines with FWHM $\gtrsim\,$500\,\kms{}, and its {geometrical} size,
{of sub-parsec scales}, is typically determined through reverberation mapping, primarily using the H$\beta$ line in nearby AGNs \citep{ Blandford1982REVERBERATIONQUASARS, GaskellLineGalaxies, Kaspi2000ReverberationNuclei, Bentz2009THEAGNs, Grier2012ReverberationGalaxies, Du2018Mapping, Cho_etal_2021ApJ}. 
The BLR is characterized by a metal-rich environment, with \feii{} being one of the most prominent contributors to the observed emission line spectrum, spanning from the ultraviolet (UV) to the near-infrared (NIR). However, it does not universally exhibit high metallicity; both early and recent studies have shown that \feii{} emission can be consistent with solar abundances in some AGN. At the same time, photoionization modeling indicates that super-solar metallicities are often required to reproduce the strongest \feii{} emitters \citep{Negrete2012Broad-lineRedshift, Panda2019TheOrientation, 2021ApJ...910..115S, Karla_2022, 2024Physi...6..216M, Floris_etal_2025}.

The low ionization gas responsible for the emission of \feii{} is predicted to be located in the outer part of the BLR \citep{Matsuoka2008LowIonizationLines, Panda2018ModelingPlane, Panda_2021_LIL_emission, Martinez-Aldama_etal_2021}. Iron emission has been studied for decades through observations and photoionization modeling \citep{Sigut1998, Sigut2003, Bruhweiler2008Modeling1, Rudy2000The1, RodriguezArdila2002, Kovacevic2010AnalysisSpectra, Marinello2016, Panda2019TheOrientation, Kovacevic-Dojcinovic_2025, Pandey_etal_2025ApJS, Floris_etal_2025}. It is the result of over 344,000 transitions, which makes the study of the emission significantly complex \citep{Verner1999NUMERICALSPECTRA}.  Historically, the \feii{} spectrum has been extensively explored, primarily through UV-optical observations. Based on these predictions, it is estimated that the \feii{} emission plays a pivotal role in cooling the BLR gas, accounting for approximately 25\% of the total gas energy budget \citep{Wills1985BroadEmission, Marinello2016}. In the last two decades, efforts have been made to expand the \feii{} studies into the NIR region \citep{Sigut2004TheoreticalRegions, Garcia-Rissmann2012AGALAXIES, sab, diasdossantos24}. 

\citet{Boroson1992TheObjects} were pioneers to establish the empirical optical \feii{} template from the prototypical AGN \feii{} emitter, I~Zw~1. They created an empirical template by removing all emission lines different from those of \feii{}, a method that continues to be used \citep{Shen2010, Shen2014TheOverview, Rakshit2017A12, Panda_etal_2024ApJS}. In the following decades, with the improvement of observation and instrument facilities, new \feii{} templates were made available and used in statistical estimation of optical AGNs properties 
\citep{Skrutskie2006The2MASS, Kovacevic2010AnalysisSpectra, dong2010prevalence, Kovacevic-Dojcinovic2015TheAGNs, Park_etal_2022ApJS}. The optical \feii{} template is commonly utilized to remove the \feii{} emission from the AGN spectrum, and mainly to estimate the \feii{} strength (namely \rfeop{}), defined as the ratio of the flux from \feii{} multiplets centered at 4570\,\AA{} (between 4434-4684 \AA) to that of the broad \hb{} component. Subsequent studies have shown that the \rfeop{} is linked to the systematic analysis of the phenomenology and evolution of AGNs through the quasar main sequence over time \citep{Sulentic2000PhenomenologyNuclei, Shen2014, Marziani2018AQuasars, Martinez-Aldama_etal_2021, Panda_2024FrASS}.


From a theoretical standpoint, photoionization codes incorporate mechanisms such as continuum fluorescence, collisional excitation, and self-fluorescence to investigate the nature of the \feii\ emission. However, they have encountered difficulties in accurately reproducing the observed UV$-$optical \feii{} spectrum 
\citep{Wills1985BroadEmission, Joly1991, Sigut1998, Pandey_etal_2025ApJS}. Advances in models have shown that the incorporation of microturbulence into the gas resulted in an enhancement of the random motion of particles, including \feii{} ions. As a result, it leads to a more efficient absorption of ionizing photons, thereby enhancing the emission of \feii{} \citep{Veron-Cetty2004TheIZw1, baldwin2004, Bruhweiler2008Modeling1, Panda2018ModelingPlane}. 
This indicates that microturbulence acts as a phenomenological parameter that boosts the \feii{} emission, i.e., increasing microturbulence can lead to \feii{} strengths comparable to those obtained with higher metallicities \citep{Panda_2021_CaFe-II}. In this sense, microturbulence can ``mimic'' or partially compensate for metallicity effects in the predicted observables of the models, without altering the intrinsic metal abundance of the BLR gas \citep{Panda_2021_CaFe-II}.

The study of NIR \feii{} emission in AGNs with the theoretical predictions of Lyman-$\alpha$ fluorescence mechanism began with \citet{Sigut1998} and \citet{Verner1999NUMERICALSPECTRA}. Shortly thereafter, the observational discovery of the \feii{} 1\,$\mu$m lines in Seyfert galaxies was reported by \citet{Rudy2000The1}. Subsequent results on the theoretical models emphasized the critical role of Lyman-$\alpha$ fluorescence in the enhancement of specific \feii{} transitions, particularly the $\lambda$\,9200 feature \citep{Sigut2003, Sigut2004TheoreticalRegions}. The observational confirmation of the $\lambda$\,9200 bump in the spectra of Type~I AGNs studied by \citet{RodriguezArdila2002}, and later reinforced by \citet{Garcia-Rissmann2012AGALAXIES, Marinello2016, Marinello2020Panchromatic1092}, pointed out the importance of the Lyman-$\alpha$ fluorescence in leading the formation of the 1\,$\mu$m lines ($\lambda$\,9997, $\lambda$\,10502, $\lambda$\,10863, and $\lambda$\,11127), commonly grouped as the \feii{} 1\,$\mu$m lines.
Moreover, \citet{Marinello2016} showed a correlation between optical and NIR \feii{} emission, suggesting that both likely share Lyman-$\alpha$ fluorescence as a common excitation channel.

Traditionally, studies on \feii{} in the UV and optical domains have dominated the landscape for many years \citet{vestergaard2001empirical, baldwin2004, Bruhweiler2008Modeling1}. However, simultaneous modeling of \feii{} in both the optical and NIR regimes has only been conducted more recently. For instance, \citet{sab, diasdossantos24} showed that it is possible to consistently replicate optical and NIR \feii{} emissions simultaneously, although the exploration was conducted within a limited parameter space. Despite advances in the field, these models still present a challenge that requires relatively extreme physical conditions to be explained. 
 

Despite several studies pointing to the potential role of Lyman-$\alpha$ fluorescence in boosting \feii{} emission, its efficiency in simultaneously enhancing both spectral regions has not been fully explored. One obstacle lies in the complex dependence of fluorescence on gas density, metallicity, and the structure of the ionizing continuum. Furthermore, low-ionization lines such as the Ca\,{\sc ii} triplet ($\lambda$$\lambda$$\lambda$8498, 8542, and 8662 $-$ hereafter CaT) and O\,{\sc i}$\lambda$8446 (hereafter O\,{\sc i}) are often observed alongside \feii{}, suggesting a shared origin and physical environment. Previous studies have shown that these lines can serve as reliable proxies for tracing the properties of the \feii{}-emitting gas \citep[e.g.][]{LoliMartinez-Aldama2015, Martinez-Aldama_etal_2021, Panda_2021_CaFe-II}, providing valuable additional constraints on the modeling.

In this study, we investigate the physical conditions responsible for the optical and NIR \feii{} emission in I~Zw~1 by testing three updated atomic datasets for the \feii{} ion \citep{Bautista_2015, tayal2018, smith19}, in addition to the widely used \citet{Verner1999NUMERICALSPECTRA} dataset, as implemented in the photoionization code \cloudy{}. These three atomic datasets were previously analyzed by \citet{Sarkar2021ImprovedSets} in the UV-optical regime. In contrast, in this present work, we are exploring for the first time these datasets in a simultaneous optical+NIR framework.

Our analysis focuses on I~Zw~1 as a representative source due to its extensively studied \feii{} emission and template-established spectrum make it an ideal reference source. Moreover, we evaluate how Lyman-$\alpha$ fluorescence contributes to optical and NIR \feii{} lines under different physical conditions, particularly varying hydrogen density and metallicity. The study is extended to include other low-ionization tracers such as the CaT triplet and O\,{\sc i} lines, enabling a broader optical-to-NIR characterization of the emitting gas. To this end, we simulate the optical (4434$-$4684~\rm\AA) and NIR (8000$-$11600~\rm\AA) spectral regions and compare model predictions with observational properties of I~Zw~1, including the flux strengths of the \feii{} 1\,$\mu$m lines relative to nearby H\,{\sc i} lines \citep{RodriguezArdila2002}.

The paper is organized as follows: Section~\ref{sec:1} presents the I~Zw\,1 data from previous observational studies. Section~\ref{sec:2} describes the modeling setup used, incorporating the photoionization code \cloudy{} \citep{CLOUDY_2023RMxAA, CLOUDY_2023RNAAS}, and the new \feii{} datasets therein. Section~\ref{sec:3} presents the results and key findings from this study. We discuss the relevant issues and possible scenarios in connection with his work in Section~\ref{sec:4} and provide a summary of our work in Section~\ref{sec:5}.

\section{PAST STUDIES AND DATA} \label{sec:1}

\subsection{I~Zw\,1 -- the prototypical \feii{} emitter}

I~Zwicky~1 (I~Zw~1) is classified as Narrow Line Seyfert 1 (NLS1) \citep{osterbrock1977spectrophotometry} located at z = 0.061, and is considered as a strong \feii{} emitter \citep{Wills1985BroadEmission, Panda2020OpticalModelling}. The spectrum exhibits overlapping and blended \feii{} features, spanning from UV to NIR spectrum. Because of this, it has been the subject of extensive studies \citep{sargent, OSTERBROCK1985TheGalaxies, Laor1997TheResults, RodriguezArdila2000VisibleGalaxies, Veron-Cetty2004TheIZw1, Kovacevic2010AnalysisSpectra, Garcia-Rissmann2012AGALAXIES, RichardsonInterpretingSpectra, Panda_2021_CaFe-II}. 

Historically, optical and UV templates have been created based on the \feii{} I~Zw~1 spectra, contributing to examining the broad emission lines, and to quantifying the optical \feii{} emission in large AGN Type-I samples \citep{vestergaard2001empirical, tsuzuki2006fe, salviander2007black, dong2010prevalence, Kovacevic2010AnalysisSpectra, Shen2011, Panda_etal_2024ApJS}. Advances in the NIR \feii{} model were made by \citet{Garcia-Rissmann2012AGALAXIES} through the semi-empirical \feii{} template using observations of I~Zw~1 and theoretical models \citep{Sigut2003, Sigut2004TheoreticalRegions}.  In particular, the template is important for the study of the four \feii{} lines located at 1\,$\mu$m region \citep{RichardsonInterpretingSpectra, Marinello2016, diasdossantos24}, which are of key interest to this work.

\subsection{Archival Data}

Our model constraints are based on the observational I~Zw\,1 optical and NIR data. The NIR spectra were observed using the 3.2\,m IRTF telescope (NASA Infrared Telescope Facility) at Mauna Kea, Hawaii, USA in 2004 \citep{Garcia-Rissmann2012AGALAXIES}.
The SpeX spectrograph was in cross dispersion mode (SXD), covering a wavelength range of 0.8\,$-$\,2.4\,$\mu$m, with a spectral resolution of $\sim$1300. The optical counterpart was obtained at the CASLEO Observatory in Argentina, using the REOSC spectrograph in long slit mode covering the range 3500\,$-$\,6800\,\AA.
Detailed observational and reduction information can be found in \citet{Garcia-Rissmann2012AGALAXIES} (NIR), and \citet{Marinello2016} (optical).

An additional near-infrared spectrum was incorporated in this work to complement our photoionization modeling. The spectrum presented in \citealt{diasdossantos24} was preserved throughout the current analysis, as it provides the observational basis for the Fe\,{\sc ii} measurements used in the previous modeling framework. This is particularly important for enabling a direct comparison between the predictions obtained with the updated \feii{} atomic models implemented in our {\sc cloudy} simulations and the observational constraints adopted in \cite{diasdossantos24}. For this reason, the \feii{} measurements from that work are retained in all main results presented here.

The new near-infrared spectrum of I\,Zw\,1 was observed on 2024 September 9 using the SpeX spectrograph mounted on the NASA Infrared Telescope Facility (IRTF). The observations were performed in ShortXD mode with a slit of 0.5$\times$15 arcsec, providing a spectral resolving power of $R \sim 1200$. The total on-source integration time was 2160\,s, with individual exposures of $\sim$180\,s. The observations were conducted at an airmass of $\sim$1.01, under stable atmospheric conditions.

The data reduction, spectral extraction, and wavelength calibration were carried out using the {\sc SpeXTool} package V4.1 \citep{Vacca2004Spextool:Spectrograph}, following standard procedures. Telluric correction and flux calibration were performed using an A0\,V standard star observed close in airmass to the science target, employing the {\sc xtellcor} routine. The individual spectral orders were subsequently combined into a final one-dimensional spectrum covering the $\sim$0.9--2.4\,$\mu$m range. The final spectrum includes an associated error vector that accounts for uncertainties propagated during the reduction and extraction processes. Finally, the spectrum was corrected for redshift and Galactic extinction using standard prescriptions.

To further minimize the number of free parameters in our simulations (described in section~\ref{modelingstup}), we adopted several observational parameters based on previous studies. We used the BLR luminosity at 5100~\AA{}, $L_{5100}$ = 3.19 $\times$ 10$^{44}$ erg s$^{-1}$ of I~Zw~1 from \citet{Kaspi2000ReverberationNuclei}, whose observation period closely matches that of the optical and NIR spectra used in this work. 
In addition, we used the BLR radius ${R_{\rm BLR}}$ = 37.2~light-days for I~Zw~1, obtained by reverberation mapping based on H$\beta$ emission \citep{Huang2019ReverberationMass}. Moreover, we utilize the spectral energy distribution (SED) of I~Zw~1 from \citet{Panda2020OpticalModelling}, which is hosted on the GitHub repository\footnote{https://github.com/Swayamtrupta/CaT-FeII-emission}. The SED was included as the ionization continuum in the \cloudy{} input files for the simulations that have provided meaningful constraints on the optical \feii{} and NIR CaT in previous studies \citep{Panda2020OpticalModelling, Panda_2021_CaFe-II}.


\subsection{Optical and NIR \feii{} intensity estimation}\label{estimations}

As we are interested in the strengths of \feii{} both in the optical and NIR regions relative to the nearest hydrogen lines, H$\beta$ and Pa$\beta$, we first calculate their line fluxes. Using optical and NIR spectra and a template of I~Zw~1, we derived key \feii{} properties such as intensities, FWHMs, and fluxes. Moreover, we derive the optical \feii{} and NIR intensities by their line flux ratios to the nearest hydrogen line. All these properties were taken from our previous work  \citep{diasdossantos24}. For the optical region, the \feii{} derived value was the \(R_{\rm 4570} = 1.62 \pm 0.06\) \citep{diasdossantos24}, which is aligned with previous studies, with values around 1.47 \citep{Sulentic2000THE1416129}. 

The \feii{} NIR intensity, $R_{1\mu\,{\rm m}}$, is defined as the ratio of the combined fluxes of the four brightest NIR isolated \feii{} lines at the wavelengths $\lambda$9997, $\lambda$10502, $\lambda$10863, and $\lambda$11127 to the flux of Pa$\beta$ broad component \citep{RodriguezArdila2002}. We estimate the flux of the broad Pa$\gamma$ component, since Pa$\beta$ lies in a region strongly affected by telluric residuals. Thus, we estimate $R_{1\mu,{\rm m}}$ by rescaling the Pa$\gamma$ flux to the Pa$\beta$ flux. However, our approach does not rely on the theoretical Pa$\beta$/Pa$\gamma$ ratio (0.8531) \citep{Marinello2020Panchromatic1092}, as our previous work identified issues with its use (see \citealt{diasdossantos24} for more details). 
Instead, we investigated how the Pa$\gamma$/Pa$\beta$ ratio is evaluated under various gas conditions using \cloudy{} simulations (see Appendix~\ref{apendicea} and \citealt{diasdossantos24} for details). To this purpose, we assumed a fixed column density of 10$^{24}$~cm$^{-2}$ for each dataset described in Section~\ref{sec:2} and their corresponding microturbulence cases. 

In Figure~\ref{apendicea}, we show the dependence of the Pa$\beta$/Pa$\gamma$ ratio on the gas density, considering the metal content range assumed in this work, using the \feii{} datasets described in \citet{smith19, tayal2018}. The $R_{1\mu\,{\rm m}}$ varies between $\sim$0.9 and $\sim$2.4 in the simulations, with slight variations among the different cases. To estimate this interval, $R_{1\mu,{\rm m}}$ was rescaled to the observed Pa$\gamma$ flux using the simulated Pa$\beta$/Pa$\gamma$ ratio, applying both the minimum and maximum values. For example, when micro-turbulence is off with the \citet{Verner1999NUMERICALSPECTRA} dataset, the $R_{1\mu\,{\rm m}}$ varies between \(0.46\pm0.47\) and \(1.48\pm0.15\) \citep{diasdossantos24} (see Table~\ref{tab:feii_comparison}).

The fitting of the emission lines, continuum, and \feii{} template for the additional I\,Zw\,1 spectrum was performed following the procedure described in \citet{diasdossantos24}.  The fit obtained for this new spectrum is shown in Appendix~\ref{new_spec}, while the corresponding measured line properties are listed in Table~\ref{tab:feii_comparison}. To evaluate our conclusions using an independent spectrum and an independent Fe\,{\sc ii} measurement, we also compare in  section~\ref{sec:3} the main results obtained using the measurements from \citet{diasdossantos24} with those derived from the new I\,Zw\,1 spectrum. Following the same procedure adopted in \citet{diasdossantos24}, we find that the \rfenir{} values derived from the new spectrum range between 0.35 and 0.85, lower than the previous estimate (0.64--1.55), with partial overlap within the uncertainties (see Table~\ref{tab:feii_comparison}).

\section{PHOTOIONIZATION MODELING} \label{sec:2}

The \cloudy{} code is widely used for simulating the physical conditions in a gas by solving the radiative transfer equations through escape probability formalism, either under Local Thermodynamic Equilibrium (LTE) or non-LTE scenarios, making it a versatile tool for wide astrophysical environments \citep{CLOUDY_2023RMxAA, CLOUDY_2023RNAAS}. The version (v23.01) introduces three additional \feii{} atomic datasets, significantly improving the modeling of \feii{} emission lines and providing more accurate predictions of the \feii{} pseudo-continuum. Furthermore, \cloudy{} provides an extensive atomic database relative to other codes. These three new datasets allow us to evaluate the state-of-the-art \feii{} atomic models to recover the line intensities around 1\,$\mu$m region.
A broader overview of these new datasets can be found in \citet{Sarkar2021ImprovedSets}. Below, we present a summary of the most significant characteristics of each of these new datasets that are pertinent to our research.


\subsection{Overview of the \feii{} datasets used in this work}

Besides \citet{Verner1999NUMERICALSPECTRA}, we applied three recent Fe\,{\sc ii} atomic datasets, \citet{Bautista_2015}, \citet{tayal2018} and \citet{smith19} to our spectral synthesis models.

\citet{Bautista_2015} present an ion with 159 levels up to $\sim$\,11.56\,eV, producing 628 emission lines. In particular, the \feii{} ion comprises 52 lowest metastable levels, which account for essentially forbidden lines in the optical and IR spectral regions. This allows the atomic model to be used to explore mainly forbidden [\feii{}] lines in optical and NIR spectral regions. Although we are mostly interested in the 4000$-$12000~\AA\ spectral region, it is important to point out that \citet{Bautista_2015} reports no transition between 2000$-$3000~\AA.

\citet{tayal2018} \feii{} model contains 340 energy levels, with energies up to $\sim$\,16.6\,eV, producing 57635 emission lines. This dataset allows a more detailed treatment of the \feii{} spectra from astrophysical sources, which is particularly interesting for our study. We point out that the \citet{tayal2018} model does not necessarily include the \citet{Bautista_2015} energy levels.

\citet{smith19} report 20 new configurations, a 6069 level atomic structure model, including 716 energy levels that produce 255974 emission lines with the highest energy level of $\sim$\,26.4\,eV. This dataset is large compared to \citet{tayal2018} and includes more relevant forbidden and permitted transitions. One of the benefits of \citet{smith19} is that the electrons can achieve higher energy levels ($\lesssim$ 13\,eV), which is important for \feii{} NIR emission.  

\citet{Garcia-Rissmann2012AGALAXIES} (hereafter, GR) developed a semi-empirical \feii{} template from the prototypical \feii{} emitter I~Zw\,1. This template has been used to model the \feii{} emission in several AGNs,  specifically, to estimate the \feii{} NIR ratio in these \feii{}-rich AGNs. We compare the modeled pseudo-continua with the GR template to choose the ideal dataset that best reproduces the \feii{} line intensities and ratios in our study.


Although \citet{tayal2018} has a similar number of levels (340) compared to the previous default \feii{} dataset in \cloudy{} \citep{Verner1999NUMERICALSPECTRA}, which has 371 levels, we observed salient differences between them \citep{diasdossantos24}. Additionally, since \citet{smith19} extends to higher energy levels (up to 13 eV), which are crucial for NIR \feii{} emission \citep{RodriguezArdila2002, Marinello2016, Marinello2020Panchromatic1092}, we conducted our analysis focused on \citet{tayal2018} and \citet{smith19} datasets.


\subsection{Modeling setup}\label{modelingstup}

We perform a suite of \cloudy{} models 
{to evaluate the \feii\ datasets that best reproduces the observed \feii\ strength in I\,Zw\,1.} To minimize the free parameters in our simulations, (i) we adopted the spectral energy distribution (SED) for I~Zw~1 from \citet{Panda2020OpticalModelling} as the input radiation field in our simulations, (ii) we use the continuum luminosity at 5100~\AA{} of I~Zw~1 from \citet{Kaspi2000ReverberationNuclei}, whose observation date closely matches that of the optical and NIR spectra used in this work. The luminosity at 5100~\AA{} adopted is \(L_{5100} = 3.19 \times 10^{44} \text{ erg s}^{-1}\) \citep{Kaspi2000ReverberationNuclei}, and (iii)  the BLR radius employed is $R_{\rm BLR}$\,=\,37.2\,light-days, measured from reverberation mapping by \citet{Huang2019ReverberationMass}. All models share the same input as SED, continuum luminosity, and BLR radius.

Thus, we limit the free parameters to 3, namely the cloud column density (N$_{\rm H}$), the cloud local hydrogen density (n$_{\rm H}$), and the metal content of the BLR ($Z$). 
We created a set of 609 \cloudy{} models by varying the local hydrogen density in the interval 10$^{7}\,-\,$10$^{14}$\,cm$^{-3}$. The metallicity varied between 0.1$Z_{\odot}$ and 10\,$Z_{\odot}$  using the \textit{GASS10} module \citep{Grevesse2010TheSun}. The column density was set to 10$^{24}$\,cm$^{-2}$. Lower column density regimes (N$_{\rm H}\,=$\,10$^{22}$\,cm$^{-2}$, 10$^{23}$\,cm$^{-2}$) were already explored by \citet{sab}. The physical conditions incorporated were motivated by previous studies for strong \feii{} emitting AGNs \citep{Panda2018ModelingPlane,Panda2019TheOrientation, Panda2020OpticalModelling,Sarkar2021ImprovedSets,Sniegowska,Karla_2022}, where a more complete description of the assumed range of values can be found. 

In addition, we explored the impact of microturbulence by testing different velocities: 0 \kms{}, 10 \kms{}, and 100 \kms{}. These values were adopted to evaluate how microturbulent broadening affects the predicted \feii{} emission, specifically the \rfeop{} and \rfenir{} ratios. A detailed discussion of the outcomes is provided in the following sections.

\begin{figure*}
\centering
\includegraphics[width=\textwidth]{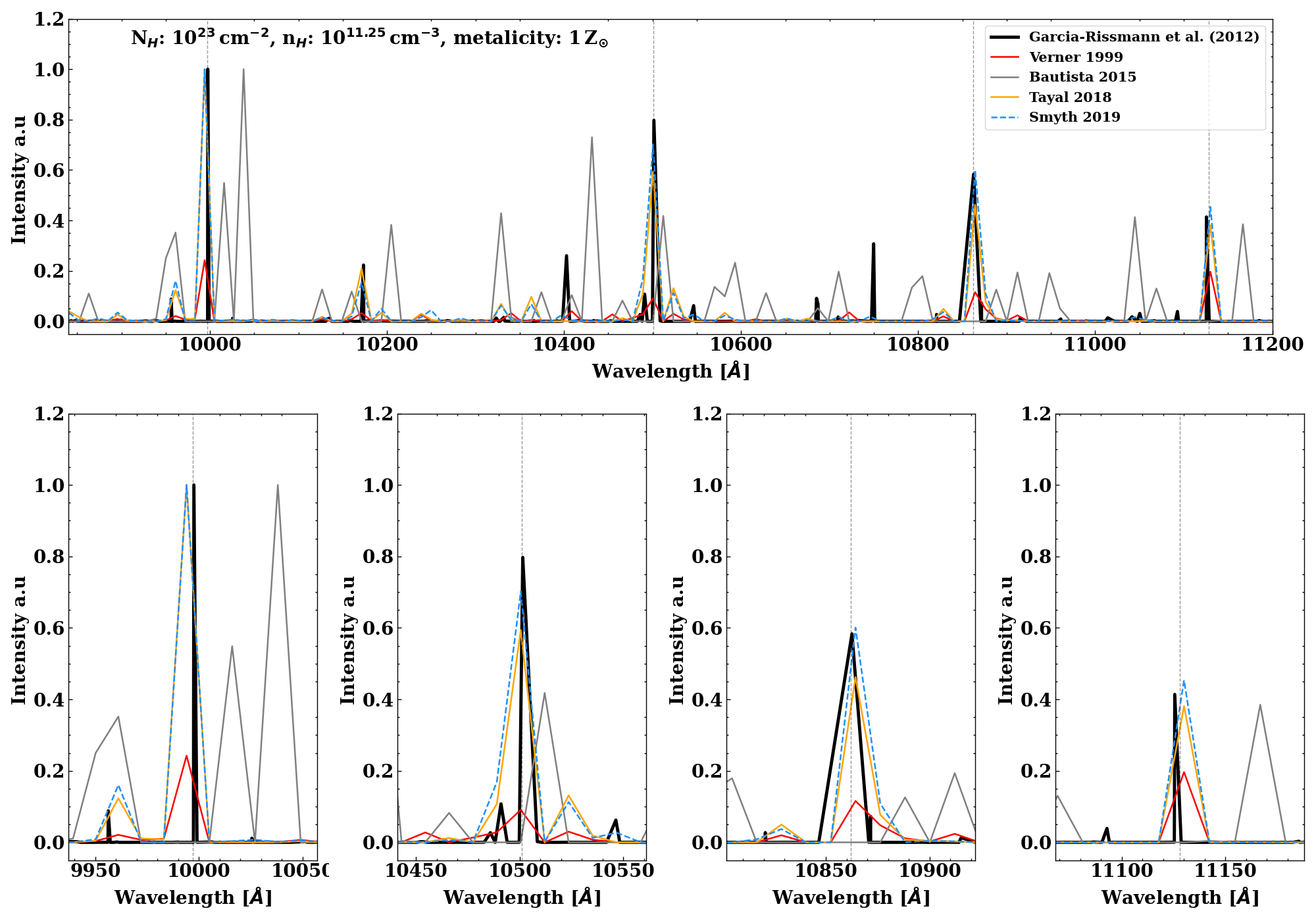}
\caption{Upper panel: Predicted pseudo-continuum for NIR \feii{} emission using a representative case of physical conditions, n$_{\rm H}$=10$^{11.25}$\,cm$^{-3}$, solar metallicity and column density case 10$^{24}$\,cm$^{-2}$. The atomic databases -- \citet{Verner1999NUMERICALSPECTRA}, \citet{Bautista_2015}, \citet{tayal2018}, and \citet{smith19}, are shown respectively in red, gray, orange, and dashed blue. The \citet{Garcia-Rissmann2012AGALAXIES} template is shown in black. Lower panel: Zooming in on the \feii{} 1\,$\mu$m lines.}
\label{fig:1}
\end{figure*}

A representative \feii{} pseudo-continuum generated by a set of physical conditions is present in Figure~\ref{fig:1}. We compared the simulated spectra, \citep{Verner1999NUMERICALSPECTRA, smith19, tayal2018, Bautista_2015}, normalized to their respective intensity maxima between 9000$-$12000\,\rm\AA~with the GR template, in wavelengths and intensities of the \feii{} lines. 
We note that \citet{Verner1999NUMERICALSPECTRA} predicts an additional strong \feii{} line that is not present in the \citet{Garcia-Rissmann2012AGALAXIES} spectrum, highlighting a key difference between the two models. Regarding the \citet{Bautista_2015} dataset, it fails to predict the lines that are of interest to us. This is possibly explained by the fact that the model is primarily designed to predict forbidden transitions. Conversely, the other two datasets, \citet{smith19} and \citet{tayal2018}, predict the permitted lines that we are interested in and show better agreement with the \citet{Garcia-Rissmann2012AGALAXIES} template.
In the next section, we explore further out the analysis using the atomic datasets from \citet{smith19} and \citet{tayal2018}, which provide meaningful constraints on the physical conditions of the \feii{}-emitting gas in the BLR of this source.

\section{Analysis and Results} \label{sec:3}

Microturbulence refers to small-scale, random gas motions within the emitting regions of the BLR that are not resolved in the macroscopic flow. These turbulent motions increase line broadening and enhance the interaction between radiation and \feii{} ions, leading to more efficient absorption of ionizing photons and consequently to stronger \feii{} emission. This effect is well-established as an essential ingredient in {\sc cloudy} models of \feii{} emission in AGNs, as demonstrated in previous works for the UV and optical regimes \citep{Veron-Cetty2004TheIZw1, Bruhweiler2008Modeling1, Panda2018ModelingPlane, Panda_2021_CaFe-II, zhang2024insights}.

In particular, microturbulence can improve the optical \feii{} strength, \rfeop{}, in a way analogous to increasing the abundance of metals, effectively acting as a metallicity regulator \citep{Panda_2021_CaFe-II}. In this work, rather than revisiting its general importance, we investigate how microturbulence influences the physical conditions required to reproduce both \rfeop{} and \rfenir{} when using updated \feii{} atomic datasets and extending the analysis to the NIR regime. This approach allows us to assess whether the role of microturbulence in the NIR follows the same trends observed in the optical regime.
A detailed discussion of these results is presented in the following sections.


\subsection{Analysis without Micro-turbulence}

\begin{figure*}
    \centering
    \includegraphics[width=\textwidth]{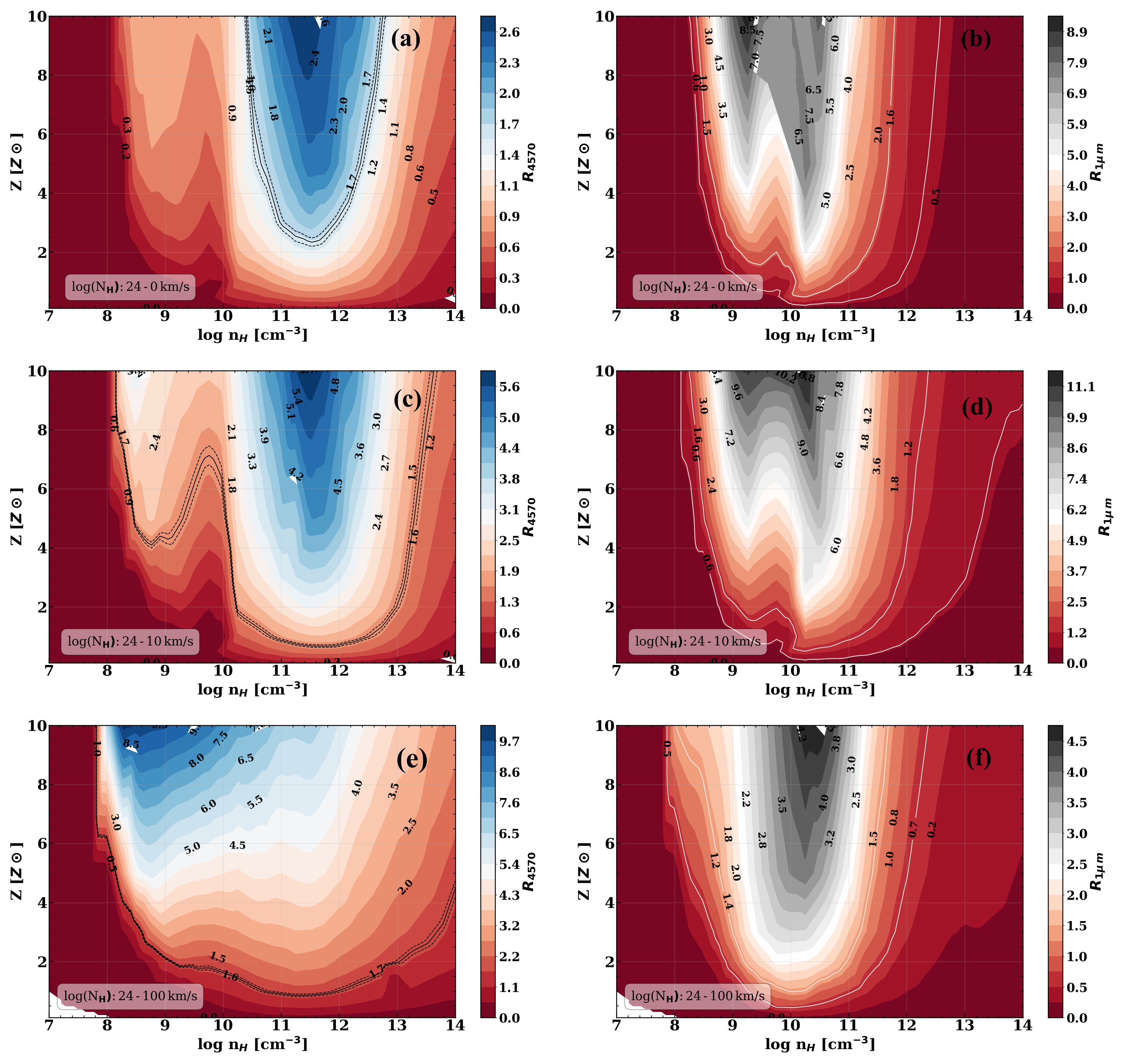}
    \caption{Parameter space, showing hydrogen density (log\,n$_{H}$) versus metallicity (Z$_{\odot}$) for models using the Smyth \feii{} dataset. The panels in the left column show the results for R$_{4570}$ while those in the right column display those of R$_{1\mu\,m}$. From top to bottom, the panels represent differences in microturbulence velocities: $v$\,=\,0\,\kms{}, $v$\,=\,10\,\kms{}, and $v$\,=\,100\,\kms{}. The white solid lines represent the range of values for Pa$\beta$ flux and \feii{} 1\,$\mu$m line intensities based on our approach (see Section~\ref{sec:2}). The solid black line and the dashed black lines, respectively, are the observed value of \rfeop{} and associated error bars.}
    \label{fig:smyth_ratio_feii}
\end{figure*}

\begin{figure*}
    \centering
    \includegraphics[width=\textwidth]{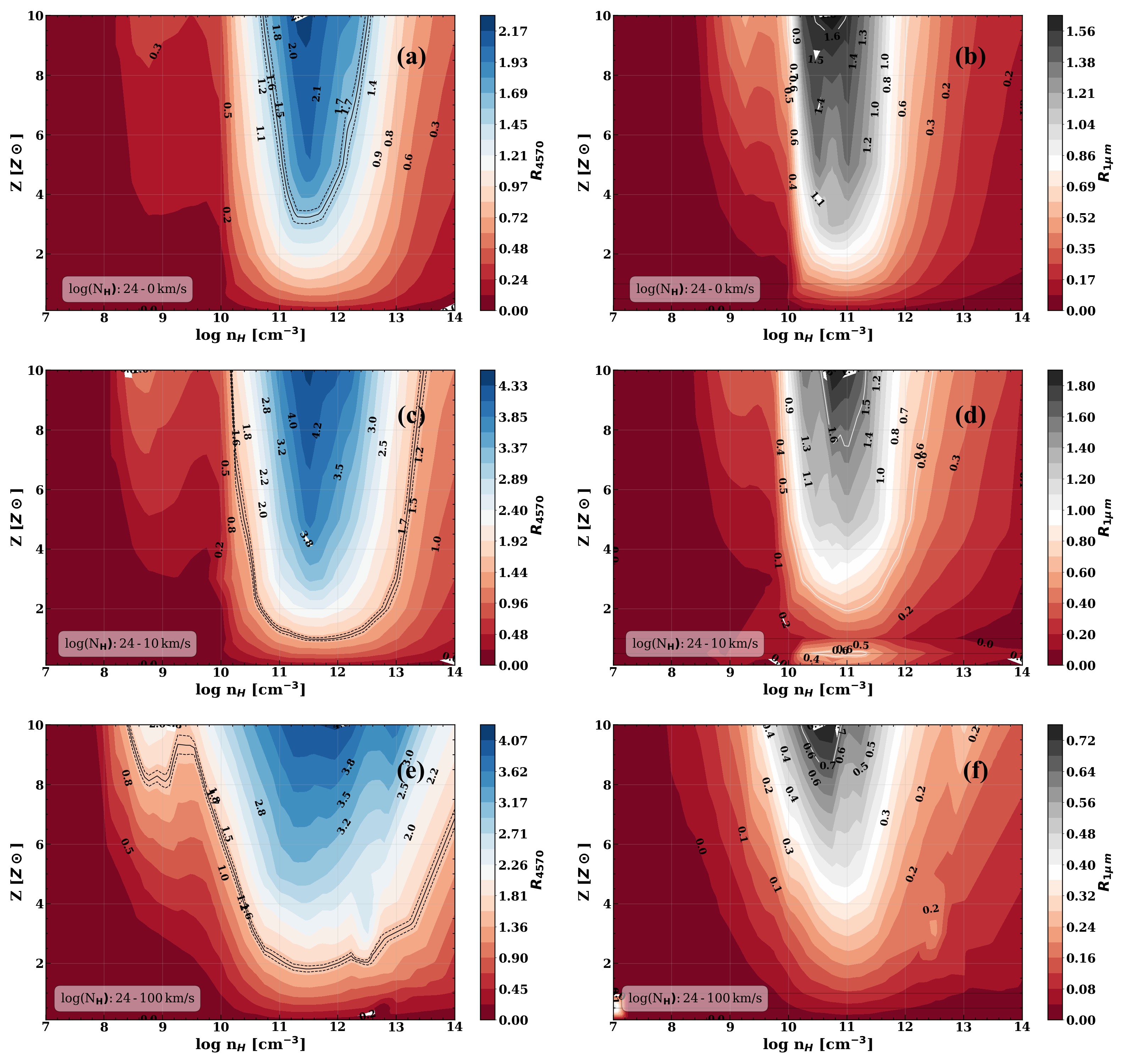}
    \caption{Same as Figure~\ref{fig:smyth_ratio_feii} but for Tayal \feii{} atomic dataset.}
    \label{fig:tayal_ratio_feii}
\end{figure*}

In Figures~\ref{fig:smyth_ratio_feii} and \ref{fig:tayal_ratio_feii}, the panels present the parameter space of metallicity versus local hydrogen density for the \citet{smith19} and the \citet{tayal2018} datasets. The colour bar represents the  \rfeop{} (left column) and \rfenir{} (right column) ratios. The top raw in both figures shows model results without microturbulence (0 kms{}).

Figures \ref{fig:smyth_ratio_feii} and \ref{fig:tayal_ratio_feii} include key observational constraints. The solid black line represents the observed value of \rfeop{} (the ratio of the \feii{} multiplet centered at 4570\,\AA{} to H$\beta$), with dashed black lines marking the associated error. The white solid lines indicate the physical range of \rfenir{}, derived by rescaling the Pa$\gamma$ flux using the minimum and maximum Pa$\beta$/Pa$\gamma$ ratios from our simulations, and combining these with the limits of the summed NIR \feii{} line fluxes. The estimates were obtained by rescaling the observed Pa$\gamma$ flux using Pa$\beta$/Pa$\gamma$ ratios predicted by \cloudy{} simulations, which were performed under a range of gas conditions, including different hydrogen densities and metallicities (see Appendix~\ref{apendicea} and \citealt{diasdossantos24}).


Our results using the \citet{smith19} \feii{} atomic model are presented in Figure~\ref{fig:smyth_ratio_feii} - panels (a) and (b)- and show that both \rfeop{} and \rfenir{} ratios can be reproduced under comparable physical conditions. The simultaneous physical conditions found include local hydrogen density ranging between 10$^{10.50}$ - 10$^{12.50}$ cm$^{-3}$ and metallicity greater than 2\,$Z_{\odot}$. In particular, the observed \rfenir{} values can be reproduced from subsolar metallicity, i.e., 0.5\,$Z_{\odot}$, within a local density range of 10$^{10.0}$ - 10$^{11.0}$ cm$^{-3}$. However, reproducing the optical intensity \feii{} requires a slightly higher metal content ($\gtrsim$ 2,$Z_{\odot}$) and a slightly different n$_{H}$ range, i.e., 10$^{11.00}$ - 10$^{12.00}$ cm$^{-3}$. Such metallicities were not obtained in previous studies that used the \citet{Verner1999NUMERICALSPECTRA} dataset \citep{diasdossantos24}. Without microturbulence, \citet{Panda_2021_CaFe-II} obtained a metallicity $\sim$\,10$-$20\, $Z_{\odot}$. Thus, our results represent a significant improvement provided by the application of the updated datasets utilized in the current study.

The results obtained from the \citet{tayal2018} dataset are presented in Figure~\ref{fig:tayal_ratio_feii} (panels (a) and (b) ). We reproduce \rfeop{} and \rfenir{} under overlapping physical conditions, with local hydrogen densities spanning from 10$^{11.25}$ to 10$^{11.50}$ cm$^{-3}$ and metallicity greater than 3\,$Z_{\odot}$.
It is noteworthy that the \rfenir{} is reproduced within a density range of 10$^{10.50}$ - 10$^{12.50}$ cm$^{-3}$, with metallicity above 2\,$Z_{\odot}$. Within a similar n$_{H}$ range, we obtained \rfeop{} from 10$^{11.25}$ to 10$^{11.75}$ cm$^{-3}$; however, this requires a higher metallicity condition, i.e.,  $Z \gtrsim$ 3\,$Z_{\odot}$. 

From the results using the \citet{smith19} and \citet{tayal2018} databases, we note that despite the \feii{} improvements, the optical results remain very comparable across different models, likely because the number of transitions remains the same among them. Nevertheless, a clear improvement is observed in the NIR counterpart with the newer models, particularly using \citet{smith19}. This is likely due to differences in their respective energy levels, as NIR \feii{} emissions require energies above $\sim$\,7\,eV, and in some cases up to $\sim$\,13\,eV. While \citet{tayal2018} extends up to approximately $\sim$\,16.6\,eV, \citet{smith19} reaches the highest energy levels, up to $\sim$\,26.4\,eV. These energy levels are critical for NIR emission lines, as the default dataset \citet{Verner1999NUMERICALSPECTRA} only reaches up to 11.6\,eV. In contrast to NIR \feii{} transitions, optical \feii{} emission requires lower energy, around $\sim$\,5\,eV \citep{RodriguezArdila2002, Marinello2016, Marinello2020Panchromatic1092}, which is effectively accounted for in all datasets.
Additionally, there is an overall agreement between the solutions provided by both datasets. To simultaneously reproduce the optical$\,-\,$NIR \feii{}, the physical conditions required are: a local hydrogen density range of 10$^{10.50}$ - 10$^{12.50}$ cm$^{-3}$, and metallicities up to 3\,$Z_{\odot}$.



\subsection{Impact of Micro-turbulence on \feii{} Emission}

Here, we use micro-turbulence velocities of $v_{turb}$ = 10 \kms{} and 100 \kms{}, in our models, as shown in Figures~\ref{fig:smyth_ratio_feii} and \ref{fig:tayal_ratio_feii} -panels c, d, e, and f - as representative cases. We observed an improvement in the \feii{} ratios for both datasets. Considering $v_{turb}$ = 10 \kms{} (see (c) and (d) panels in Figure~\ref{fig:smyth_ratio_feii} and \ref{fig:tayal_ratio_feii}), \rfeop{} and \rfenir{} ratios can be simultaneously reproduced under overlapping physical conditions, with local hydrogen densities ranging between 10$^{11.00}-$10$^{12.00}$ cm$^{-3}$ at solar metal content. 


Figure~\ref{fig:tayal_ratio_feii} ((c) and (d) panels) shows our models for microturbulence of 10\,\kms{} using \citet{tayal2018} \feii{} dataset. In general, the results are positive for reproducing \feii{} ratios simultaneously, and they are comparable to those without micro-turbulence, much like the results obtained for the \citet{smith19} dataset. In addition to that, we observe a small local hydrogen density subset of solutions, 10$^{10.50} <$ n$_{H}$ (cm$^{-3}$) $< 10^{11.25}$, where the ratios exhibit different metal conditions, solar metallicity for optical and 0.5\,$Z_{\odot}$ for NIR, respectively. The \rfeop{} and \rfenir{} ratios are simultaneously reproduced around n$_{H} \sim$ 10$^{10.75}$ cm$^{-3}$ with $Z \sim$ 2$Z_{\odot}$.
Also, we note that the optical models shown in Figure~\ref{fig:tayal_ratio_feii} provide a broader n$_{H}$ range of solutions, which can be reproduced under solar conditions.

For completeness, we also ran models with micro-turbulence at 100 \,\kms{} for \citet{smith19} and \citet{tayal2018} atomic data, as shown in the bottom panels in Figures~\ref{fig:smyth_ratio_feii} and \ref{fig:tayal_ratio_feii} (panels (e) and (f)). This threshold has been assumed from previous works \citep{Bruhweiler2008Modeling1, Panda2018ModelingPlane, Panda_WC_2019ApJ, Sarkar2021ImprovedSets}. The set of solutions for this case broadens the range of local density, reaching values as low as n$_{H} \sim$ 10$^{8}$ cm$^{-3}$ for both \feii{} datasets. It has been shown in earlier works \citep{Panda2020OpticalModelling, Panda2022ParameterizingViewpoint} that the \feii{} emitting regime under such low densities is located beyond the dust sublimation radius \citep{Nenkova2008}, and also does not produce the observed EWs. For the \citet{smith19} case, the recovered parameter space is similar to the $v_{turb}$ = 10 \kms{}, especially for the \rfenir{}. However, for the \citet{tayal2018} dataset, the resulting overlap between the observed range (in NIR) suggests strangely high metal content ($Z \gtrsim$ 8$Z_{\odot}$). This may arise due to some non-clear numerical artifacts; nevertheless, we discard the 100 \,\kms{} case (esp. for the \citet{tayal2018} dataset) from our analysis.

Despite the inclusion of microturbulence, the range of $n_{\rm H}$ values remains consistent with the models without microturbulence. For optical \feii{} emission, we found a range between 10$^{10.75}$ -- 10$^{12.25}$ cm$^{-3}$, while for NIR emission, we have solutions between 10$^{9.75}$ - 10$^{11.75}$ cm$^{-3}$. In general, the inclusion of a nominal microturbulence velocity within the BLR shows that \rfeop{} and \rfenir{} can be better reproduced, with overlapping physical conditions, than models without microturbulence. 

We identified non-simultaneous solutions for \rfeop{} and \rfenir{} at densities below $n_{\rm H}\,=\,$10$^{10}$\,cm$^{-3}$ using \citet{smith19} and \citet{tayal2018}. However, we discarded them due to consider the proximity of the optical and NIR emitting regions \citep[e.g.,][]{Marinello2016}, and their significant difference in metallicity: solar for NIR and 4\,$Z_{\odot}$ for optical; additionally, the parameter space, in this case, falls beyond the dust sublimation radius \citep{Panda2020OpticalModelling, Panda2022ParameterizingViewpoint}.
On the other hand, we only consider solutions with hydrogen densities above 10$^{10}$\,cm$^{-3}$, as these cases allow for the simultaneous reproduction of both \rfeop{} and \rfenir{} ratios. Even in cases where it is possible to reproduce each ratio individually, the required metallicity conditions remain consistent: approximately 1\,Z$_{\odot}$ for \rfeop{} and 0.5\,Z$_{\odot}$ for \rfenir{}.


Our results indicate that, independently of the \feii{} dataset used, increasing the microturbulence of the gas allows the ratios to be reproduced under lower metallicity conditions. This finding is important because it confirms previous optical studies and demonstrates that the effect remains consistent \citep{Panda_2021_CaFe-II}. The use of microturbulence reduces the problem of a higher metal content of the BLR, such as 10\,$Z_{\odot}$, which is not simple to understand physically.

Previous results from \citet{diasdossantos24}, using the \citet{Verner1999NUMERICALSPECTRA} dataset, showed that the physical conditions required to reproduce the \feii{} ratios are $n_{\rm H}$ = 10$^{10.75}$ - 10$^{11.50}$ cm$^{-3}$ and a metal content ranging from 5 $Z_{\odot}$ to 10 $Z_{\odot}$. However, the newer datasets from \citet{smith19} and \citet{tayal2018} offer a significant improvement by reproducing the \feii{} ratios at metallicity conditions that are three times lower than those needed by the \citet{Verner1999NUMERICALSPECTRA} model \citep{diasdossantos24}.


Finally, when we analyzed the emissions individually, we note a trend independent of the dataset: NIR \feii{}  reaches its maximum ratio at hydrogen densities of $10^{10.0}-10^{11.0}$ cm$^{-3}$, whereas optical \feii{} peaks at higher densities of $10^{11.0}-10^{12.0}$ cm$^{-3}$ and tends to require higher metallicity than in the NIR. This agrees with the location of the optical emission region for the \feii{} being closer than the NIR counterpart. \citep{Marinello2016}.

\subsection{Impact of updated observational constraints}

The analysis presented so far has focused on comparing our photoionization models, incorporating the new atomic datasets, with the standard \citealt{Verner1999NUMERICALSPECTRA} Fe\,{\sc ii} model, using the observational constraints derived by \citet{diasdossantos24} as the primary reference for optical–NIR Fe\,{\sc ii} emission. This approach ensures comparability with previous studies and enables a consistent assessment of the impact of the updated atomic datasets.

To expand our modelling results, we complement this analysis with an independent comparison based on a newly obtained NIR spectrum of I\,Zw\,1 (described in Section~\ref{sec:1}). This additional spectrum provides an alternative measurement of the Fe\,{\sc ii} emission in the 1\,$\mu$m region, enabling us to evaluate how variations in the observational constraints may impact the inferred physical conditions of the emitting gas. This is the first time that such an approach is explored in the literature for the NIR \feii{}.

Table~\ref{tab:feii_comparison} shows the comparison between the Fe\,{\sc ii} measurements derived from \citet{diasdossantos24} and those obtained from the new I\,Zw\,1 dataset presented in this work. While the individual line fluxes show noticeable differences between the two datasets, the resulting $R_{1\mu{\rm m}}$ values remain partially consistent within the range of uncertainties. For instance, the new spectrum yields $R_{1\mu{\rm m}}$ values between 0.35 and 0.85, which are lower than those obtained in \citet{diasdossantos24} (0.64–-1.55).

\begin{figure*}[!htb]
\includegraphics[width=\textwidth]{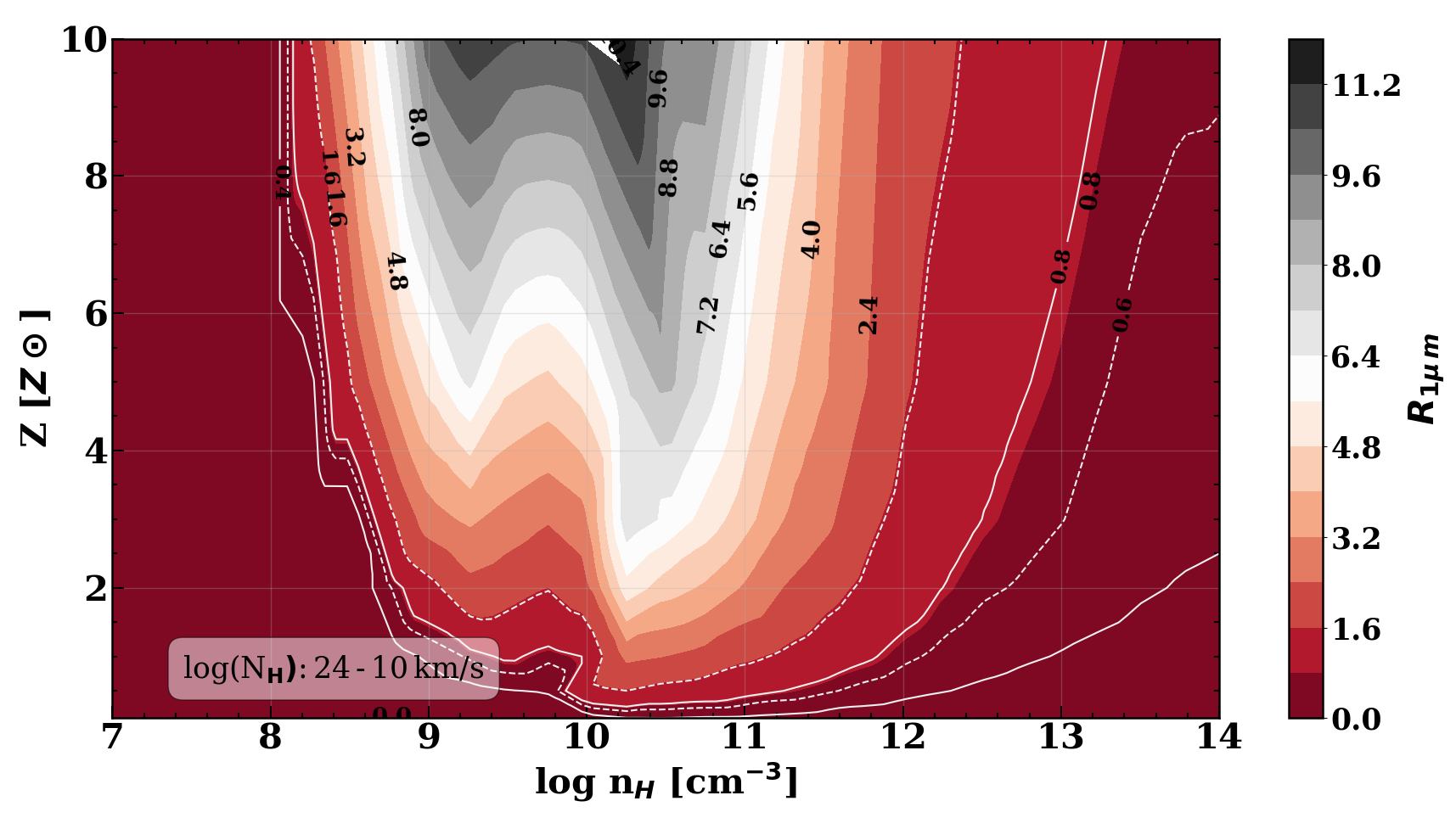}
    \caption{Parameter space, showing hydrogen density (log\,n$_{H}$) versus metallicity (Z$_{\odot}$) for models using the Smyth \feii{} dataset for the representative microturbulence velocities of $v$\,=\,10\,\kms{}. The white solid lines represent the range of values for Pa$\beta$ flux and \feii{} 1\,$\mu$m line intensities based on the new I\,Zw\,1 spectrum and the dashed-solid based on \citealt{diasdossantos24} measurement (see Section~\ref{sec:1} and Table~\ref{tab:feii_comparison}).}
\label{new_comparition}
\end{figure*}

To investigate the physical implications of this difference, we compare these observational constraints in Figure~\ref{new_comparition}, which shows the parameter space of \rfenir{} as a function of gas density and metallicity for the \citealt{smith19} setup. Since the Fe\,{\sc ii} atomic dataset of  \citealt{smith19} provides the best agreement with the observed 1\,$\mu$m emission in our models, we focus this comparison on that dataset, adopting a representative case with a microturbulent velocity of 10\,km\,s$^{-1}$. The solid contours represent the range derived from the new spectrum, while the dashed contours correspond to the previous measurements reported by \citet{diasdossantos24}. We retain the comparison with \rfenir derived by \citet{diasdossantos24} as a reference, allowing us to understand how the model behaves under observational uncertainties.

We find that the lower \rfenir{} values inferred from the new spectrum preferentially select regions of parameter space characterized by densities closer to 10$^{9.0}$\,cm$^{-3}--$10$^{10.0}$\,cm$^{-3}$, and extend towards regions of enhanced metallicity and higher density, where Fe\,{\sc ii} emission is produced more efficiently, compared to the parameter space covered by previous measurements (dashed-line).

It is important to note that, although the observational constraints are not contemporaneous, they intersect in a common region of parameter space, indicating that our main conclusions regarding the physical conditions of the emitting gas remain robust, independently of the spectral epoch in the case of I~Zw~1. This overlap defines a consistent region compatible with both constraints. Moreover, it allows us to isolate the impact of observational variations (i.e., different spectral epochs) and assess how these differences propagate into the inferred physical parameters, such as density and metallicity.

\section{Discussions}\label{sec:4}
 
\subsection{Reinforcing the paradigm of optical$-$NIR \feii{} correlation via modeling}

\begin{figure*}[!htb]
\includegraphics[width=\textwidth]{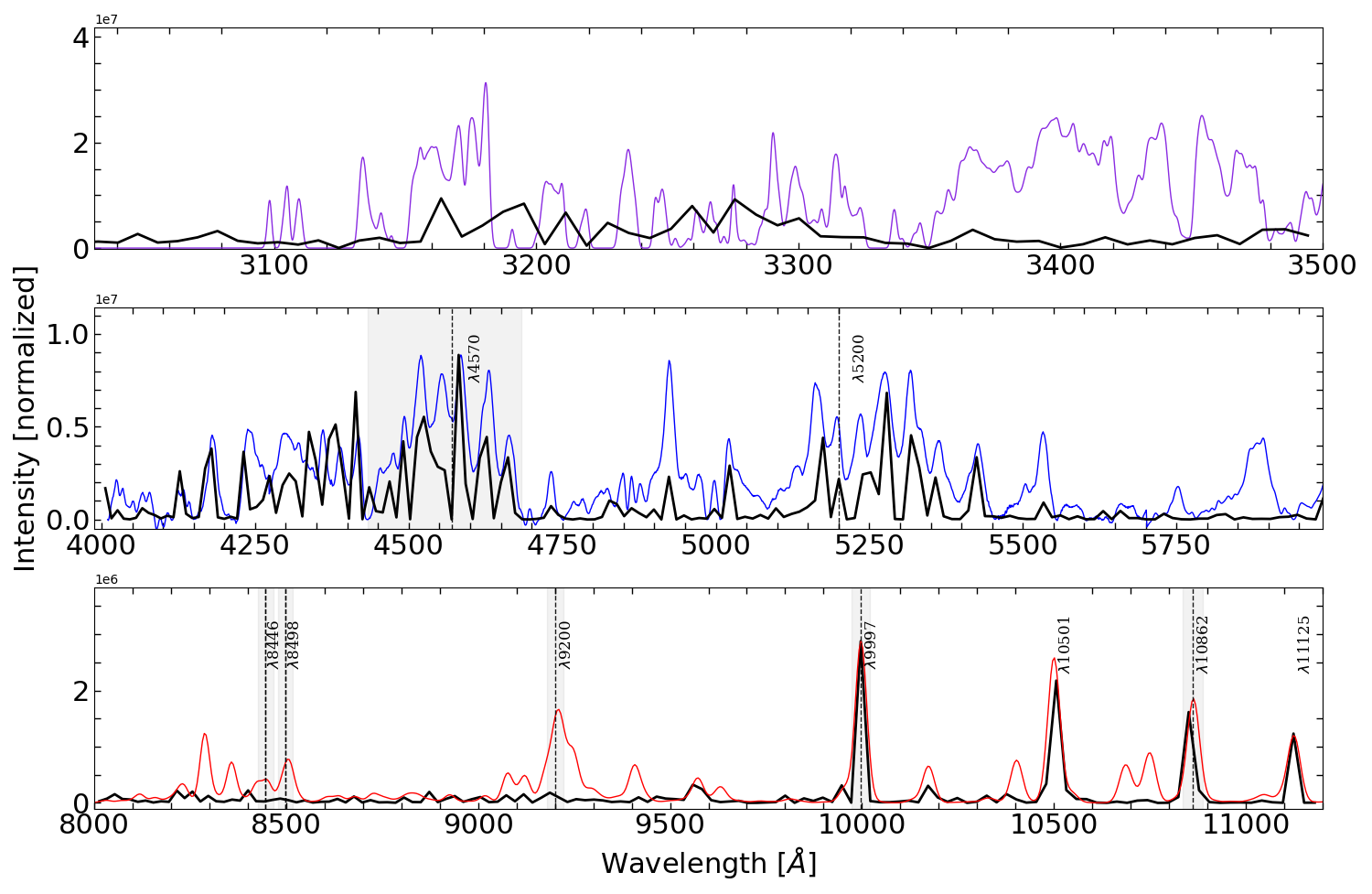}
\caption{The UV-optical-NIR \feii{} emission spectrum from a constant density (n$_{\rm H}$=10$^{10.75}$\,cm$^{-3}$) single cloud with a column density of 10$^{24}$\,cm$^{-2}$, assuming solar abundances. In purple, blue, and red, respectively, UV \citep{Veron-Cetty2004TheIZw1}, optical \citet{Boroson1992TheObjects}, and NIR \feii{} \citep{Garcia-Rissmann2012AGALAXIES}. The optical and UV are naturally broadened because they are empirical templates. The NIR was broadened by 1107\,$\pm$\,507\,\kms{} based on $\lambda\,10502$ FWHM. The \feii{} optical blue and red bumps, centered at $\lambda$4570 and $\lambda$5200, are highlighted in gray, as well the 1\,$\mu$m lines ($\lambda$9997, $\lambda$10502, $\lambda$10863, $\lambda$11127) and the 9200~\AA\ bump. The spectral region around $\lambda$8446 and $\lambda$8498 encompasses the O\,{\sc i} and CaT emission lines, respectively. The gray shaded regions highlight the spectral intervals corresponding to the Fe,{\sc ii} emission lines of interest. 
The pseudo-continuum was constructed using the \feii{} dataset from \citet{smith19}.}
\label{feii_pseudo_cont}
\end{figure*}

Unlike UV-optical lines, the NIR \feii{} emission lines have the significant advantage of being isolated or semi-isolated (see Fig.~\ref{feii_pseudo_cont}), allowing for accurate modeling and determination of individual line properties.
In the NIR spectrum, the most intense \feii{} emission lines are collectively known as the 1\,$\mu$m lines, which originate from the same energy level b$^{4}$\,G $\rightarrow$ z$^{4}$\,(F, D, P)º \citep{rudy19911,RodriguezArdila2002,Marinello2016, Marinello2020Panchromatic1092}.
The decay and consequent emission of the 1\,$\mu$m lines populate the z$^{4}$\,(F, D, P)º level, whose further decay is responsible for the optical bump centered at 4570\,\AA{}. 

From an observational point of view, the \rfeop{} and \rfenir{} ratios are intrinsically correlated, suggesting a common excitation mechanism responsible for both emissions, likely dominated by Lyman-$\alpha$ fluorescence \citep{Marinello2016}. This correlation can play a key role in allowing comparative studies of \feii{} emissions across different wavelength ranges.

In this context, our results from photoionization modeling (constant density, single cloud approximation) corroborate these observational findings by demonstrating that optical and NIR \feii{} emissions can indeed be reproduced under overlapping physical conditions for I~Zw~1 and I~Zw~1-like sources. We find that both emissions can be simultaneously reproduced using a narrow range of local hydrogen density, $\sim 10^{11.0}$~cm$^{-3}$, with nearly solar metallicity assuming a small microturbulence velocity of 10\,\kms{}. This agreement between models and observations strengthens the hypothesis that the physical conditions driving both optical and NIR \feii{} emissions are closely related. Our work links theoretical and observational results, providing physical evidence of the common environment responsible for these emissions.


To further assess the impact of observational variations on these results, we keep the \rfenir{} constraint derived from \citet{diasdossantos24} as a reference and include the new Fe\,{\sc ii} values (0.35--0.85, see Fig.~\ref{new_comparition}). We then evaluate how observational measurements affect the inferred physical conditions throughout the modeling. The comparison with the new spectrum suggests that the main physical solutions remain robust, while becoming more constrained under different observational conditions (density and metal content). The shift toward lower $R_{1\mu{\rm m}}$ values is associated with densities closer to $10^{8.0}$--$10^{9.0}$~cm$^{-3}$ and slightly lower metallicities. The new Fe\,{\sc ii} measurement suggests that the inferred conditions may be less extreme than previously indicated to reproduce the emission. At higher densities (above $10^{11.75}$~cm$^{-3}$), the Fe\,{\sc ii} emission is more efficiently produced compared to the parameter space covered by the previous measurements (dashed lines, Fig.~\ref{new_comparition}). Overall, the agreement between the new and previous Fe\,{\sc ii} estimates indicates that a consistent region of parameter space satisfies both observational constraints, preserving the main findings.


Regardless of the \feii{} atomic dataset used, we find that the NIR \feii{} emission reaches its maximum strength at hydrogen densities between $10^{10.0}$ and $10^{11.0}$~cm$^{-3}$, under sub-solar metallicities (0.5$Z_{\odot}$). In contrast, the optical \feii{} emission tends to peak at slightly higher densities, between $10^{11.0}$ and $10^{12.0}$~cm$^{-3}$, and requires at least solar metallicity. Despite these differences in peak conditions, there is a broad range of overlapping physical parameters where both emissions can co-exist and be reproduced simultaneously. This supports the interpretation that optical and NIR \feii{} emissions may arise from adjacent or partially overlapping zones within the BLR, rather than from entirely distinct regions.

This behavior is consistent with previous results using the \citet{Verner1999NUMERICALSPECTRA} dataset, which also showed that the  \feii{} in the optical and NIR reaches its peak ratio at different gas densities \citep{diasdossantos24}. Notably, this distinction remains even under low column density conditions \citep[e.g., $10^{23.0}$cm$^{-2}$, ][]{sab}. 

\subsection{Impact of Lyman-$\alpha$ excitation on optical and NIR \feii{} emissions}

It has been more than two decades since Lyman-$\alpha$ fluorescence was proposed as the primary mechanism responsible for the \feii{} emission at 9200~\AA\ \citep{Sigut1998}, and a major contributor to the NIR \feii{} spectral features, later confirmed observationally \citep{RodriguezArdila2002, Marinello2016}. Lyman-$\alpha$ can excite energy levels up to 13\,eV, and through the decay process contributes to populate lower energy levels (around 5\,eV) that are responsible for optical \feii{} emission. However, several key questions remain: Does Lyman-$\alpha$ affect the optical and NIR \feii{} emissions differently? Does its contribution vary with different physical conditions? And to what extent does this mechanism contribute to the overall emission in these wavelength ranges? 
Investigating the role of Ly$\alpha$ fluorescence is, therefore, essential to interpret the origin of the optical–NIR \feii{} correlation \citep{RodriguezArdila2002, Marinello2016}.

Previous studies have established the importance of Ly$\alpha$ fluorescence, both from modeling \citep{Sigut1998} and observational evidence \citep{RodriguezArdila2002, Marinello2016}. However, none have performed a joint optical+NIR analysis that enables a direct, condition-by-condition quantification of Ly$\alpha$ pumping across wavelength regimes. \citet{Marinello2016}, for instance, suggested that Ly$\alpha$ fluorescence generally has a stronger impact on optical \feii{} than on NIR lines, but this conclusion was not derived from photoionization modeling and was not explored as a function of the gas physical conditions.

In this section, we will carry out the first attempt to quantifying how the Lyman-$\alpha$ fluorescence influences on the \feii{} emission by varying the physical conditions. Moreover, we want to establish how its relative influence on \feii{} emission varies with the gas physical conditions, and to establish how this influence changes across different atomic data sets and spectral regimes. 

For this reason, to evaluate the influence of Lyman-$\alpha$ fluorescence around 1\,$\mu$m and the \feii{} bump centered at 4570~\rm\AA, we consider two sets of simulations within our \cloudy{} setup: one with the Lyman-$\alpha$ pumping enabled (this is switched on in \cloudy{} by default) and one without it. It is important to note that this does not represent a physically realistic BLR scenario, but instead constructs diagnostic ON/OFF ratios that isolate the Ly$\alpha$ contribution and quantify how the pumping efficiency varies across physical conditions. This controlled comparison is essential to assess whether optical and NIR \feii{} emission respond to Lyman-$\alpha$ fluorescence in the same way, and whether this behavior depends on the adopted atomic dataset and gas parameters.



\begin{figure*}
\centering
  \includegraphics[width=\textwidth]{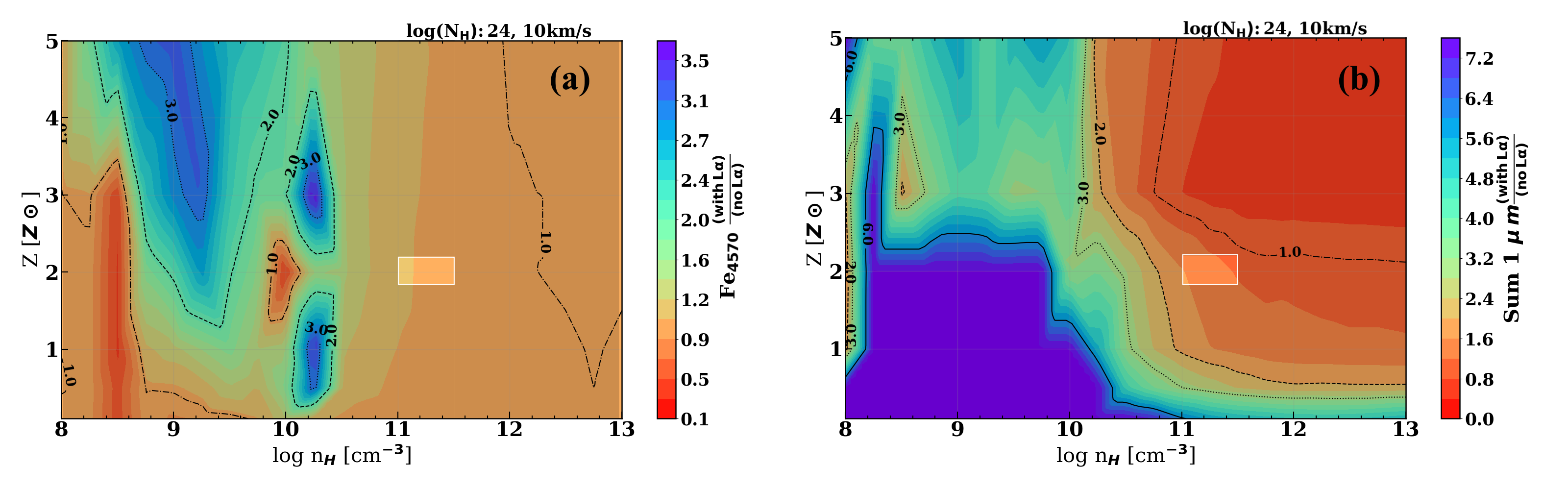}
  \caption{The contour plots show the impact of the Lyman-$\alpha$ mechanism on optical and NIR \feii{} emission along the gas physical conditions. The ratio of \feii{} emission with and without Lyman-$\alpha$ excitation for the 4570~\AA (left) and 1\,$\mu$m (right) lines is represented in color scales. The plots are a function of hydrogen density ($\log n_H$) and metallicity ($Z_{\odot}$) for a turbulence velocity of 10\,\kms{} and a column density of $10^{24}$\,cm$^{-2}$. The rectangle bolder region represents the simultaneous solutions out of the range of solutions that reproduce the \feii{} emission in both NIR and optical.}
  \label{fig:lyman-ratios}
\end{figure*}

For these models, we considered hydrogen densities ranging from 10$^{8.0}$ to 10$^{13.0}$~cm$^{-3}$, metallicities from 0.1 to 5 times the solar value, and a microturbulence velocity of 10~\kms{} for a column density, N$_{H}$\,=10$^{24.0}$~cm$^{-2}$.
By comparing the luminosity ratios from the two cases (with and without Lyman-$\alpha$ pumping) as a function of hydrogen density and metallicity, we get clues on the impact of Lyman-$\alpha$ fluorescence in the 1\,$\mu$m and 4570~\AA\ emissions. The result, shown in Figure~\ref{fig:lyman-ratios}, reveals how Lyman-$\alpha$ excitation impacts the optical and NIR spectra. It is important to note that the shaded region represents the set of physical conditions common to both emissions and required to simultaneously reproduce the observed \feii{}. The highlighted area represents the regions where the simultaneous solutions are found.

Overall, we observe that the presence of Lyman-$\alpha$ excitation leads to an increase in the \feii{} intensities in both the optical and NIR regimes, particularly notable at lower hydrogen densities and metallicities.  
The \rfeop{} value close to 1, at n$_{H}\,=\,$10$^{10.75}$~cm$^{-3}-$10$^{11.00}$~cm$^{-3}$ and solar metallicity (see Fig.~\ref{fig:lyman-ratios}), indicates that the effects with and without Lyman-$\alpha$ are relatively indistinguishable under these physical conditions for optical \feii{}. We observe similar behavior for 1\,$\mu$m lines, in densities exceeding 10$^{11.0}$ cm$^{-3}$ and in high metallicities ($>\,2\,Z_{\odot}$). This indicates that the presence of other excitation mechanisms, such as collisional excitation, is significant in these dense environments. Consequently, Lyman-$\alpha$ may not be the primary excitation mechanism under these conditions, likely responsible for the decrement in the \feii{} intensity observed at 1\,$\mu$m.

In the absence of Lyman-$\alpha$ excitation (see Figure~\ref{apendiceb}), we observe a decrement in the luminosity of \feii{} emission (Appendix~\ref{apendiceb}). Moreover, emission from the 1-micron lines is still present, indicating that collisional excitation and other mechanisms (e.g., \feii{} pumping) still contribute to the excitation processes. However, these mechanisms contribute much less significantly than Lyman-$\alpha$ \citep{Marinello2016}. From these results, we infer that when Lyman-$\alpha$ photons are abundant, they become the dominant excitation source, leading to enhancement of the emissions, as illustrated in Figure~\ref{fig:smyth_ratio_feii} (middle panel).


Although the optical and NIR results are highly correlated, the effect of Lyman-$\alpha$ on the 4570\,\AA~bump is not as significant as it is in the NIR, as shown by the comparison of Figures~\ref{fig:smyth_ratio_feii} (Lyman$-\alpha$ on) and Appendix~\ref{apendiceb} (Lyman$-\alpha$ off).
The reason for this can be attributed to several factors - efficiency of energy transfer, energy redistribution during the transitions, and non-radiative de-excitation processes, in addition to preferentially denser regions producing the optical \feii{} lines.

\begin{figure*}[!htb]
\includegraphics[width=\textwidth]{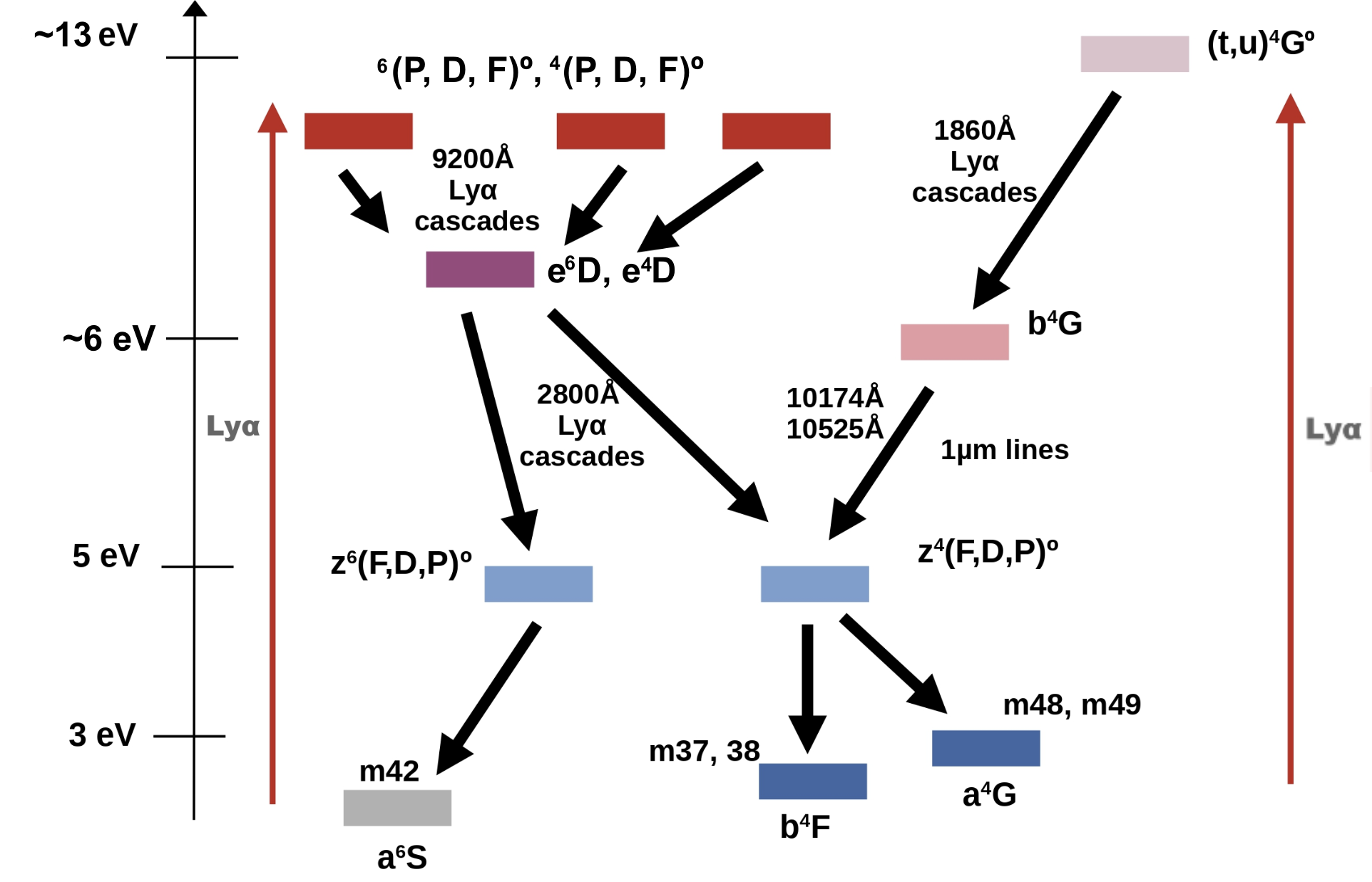}

\caption{Simplified Grotrian diagram for \feii{}, adapted from the full scheme to include only the transitions of interest in this work. The diagram highlights the role of Ly$\alpha$ fluorescence in populating high-energy levels (up to $\sim$15~eV), leading to the 9200~\rm\AA\ bump and the 1\,$\mu$m \feii{} lines ($\lambda$9997, $\lambda$10502, $\lambda$10863, $\lambda$11127), as well as subsequent cascades responsible for the optical bumps at 4570~\rm\AA ~and 5200~\rm\AA.
}

\label{gotrian}
\end{figure*}

The energy redistribution in the decay process consists of probabilities that depend on the Einstein coefficients ($A_{ki}$), based on the values available from the NIST Atomic Spectra Database (NIST)\footnote{National Institute of Standards and Technology: \url{https://physics.nist.gov/asd}} \citep{NIST_ASD}. For the transitions of interest shown in the Grotrian diagram (see Figure~\ref{gotrian}), the corresponding $A_{ki}$ coefficients indicate that optical and NIR emissions have different radiative decay probabilities across the energy levels.  
For example, the $A_{ki}$ coefficients for UV-NIR transitions are significantly higher than those for optical transitions. This suggests that most of the energy after the initial excitation by Lyman$-\alpha$ is transformed into UV and NIR photons. The cascading-down of the electrons eventually populate the upper Z$^6$ and Z$^4$ levels, which responsible for the optical emissions around H$\beta$. This may explain why Lyman$-\alpha$ does not maximize optical emissions as it does in the NIR because the upper levels from where the optical transitions originate can be easily populated by, for example, collisional excitation.


It is essential to highlight also that the 9200\,\AA~\feii{} emission provides crucial information to study the influence of Lyman$-\alpha$, and can lead to more accurate recovery of the NIR \feii{} spectrum \citep{Sigut1998,Sigut2003,RodriguezArdila2002,Marinello2016}. In this work, we note that no available atomic \feii{} dataset suitably reproduces the  9200\,\AA~emission.
The improvement of energy levels, $^{6}$(P, D, F)º and $^{4}$(P, D, F)º, responsible for the 9200\,\AA~is critical as it may have the potential to bias UV studies, particularly in the secondary Lyman$-\alpha$ cascades from the e$^{6}\,$D and e$^{4}$D group at 2800\,\AA~ (as shown in the Grotrian diagram in Figure~\ref{gotrian}), and primary transitions from the (t, u)$^{4}\,$Gº group at 1800\,\AA. The improvement of the atomic datasets is crucial for improving the accuracy of recovering both NIR and UV spectroscopic measurements. However, this task is out of the scope of this paper.

\subsection{The \feii{}$-$CaT$-$O\,{\sc i} connection}



\citet{persson} was a pioneer in proposing a positive correlation between CaT and O\,{\sc i} emissions in AGNs with strong \feii{} emissions. 
Building on this starting point, several observational and theoretical studies \citep{Joly1991, RodriguezArdila2002, Matsuoka2008LowIonizationLines, Panda2020OpticalModelling, Panda_2021_CaFe-II, Martinez-Aldama_etal_2021} have investigated the idea of using these less complex ions as a proxy to study the \feii{} gas, to simplify the spectral analysis and theoretical modeling. 

From observational results, \feii{}, CaT and O\,{\sc i} emissions show similar FWHM values \citep{Joly1991, RodriguezArdila2002, Marinello2016, Martinez-Aldama_etal_2021}, indicating that these ions share similar kinematics, and likely originate from a co-spatial region. Moreover, theoretical photoionization models \citep{jolly1987, LoliMartinez-Aldama2015, Panda2020OpticalModelling, Panda_2021_LIL_emission} show that \feii{}, O\,{\sc i}, and CaT ions are emitted in regions with similar physical conditions, such as local hydrogen density ($n_{\rm H}$), ionization parameter ($U$), and column density ($N_{\rm H}$). Those previous modeling studies have focused on physical properties derived from UV$-$optical \feii{} emissions. In contrast, our work utilizes a simultaneous $-$optical and NIR $-$ approach, providing a distinct and seldom explored analysis of the environment for \feii{} emitting gas.

\begin{figure}
\centering
  \includegraphics[width=\columnwidth]{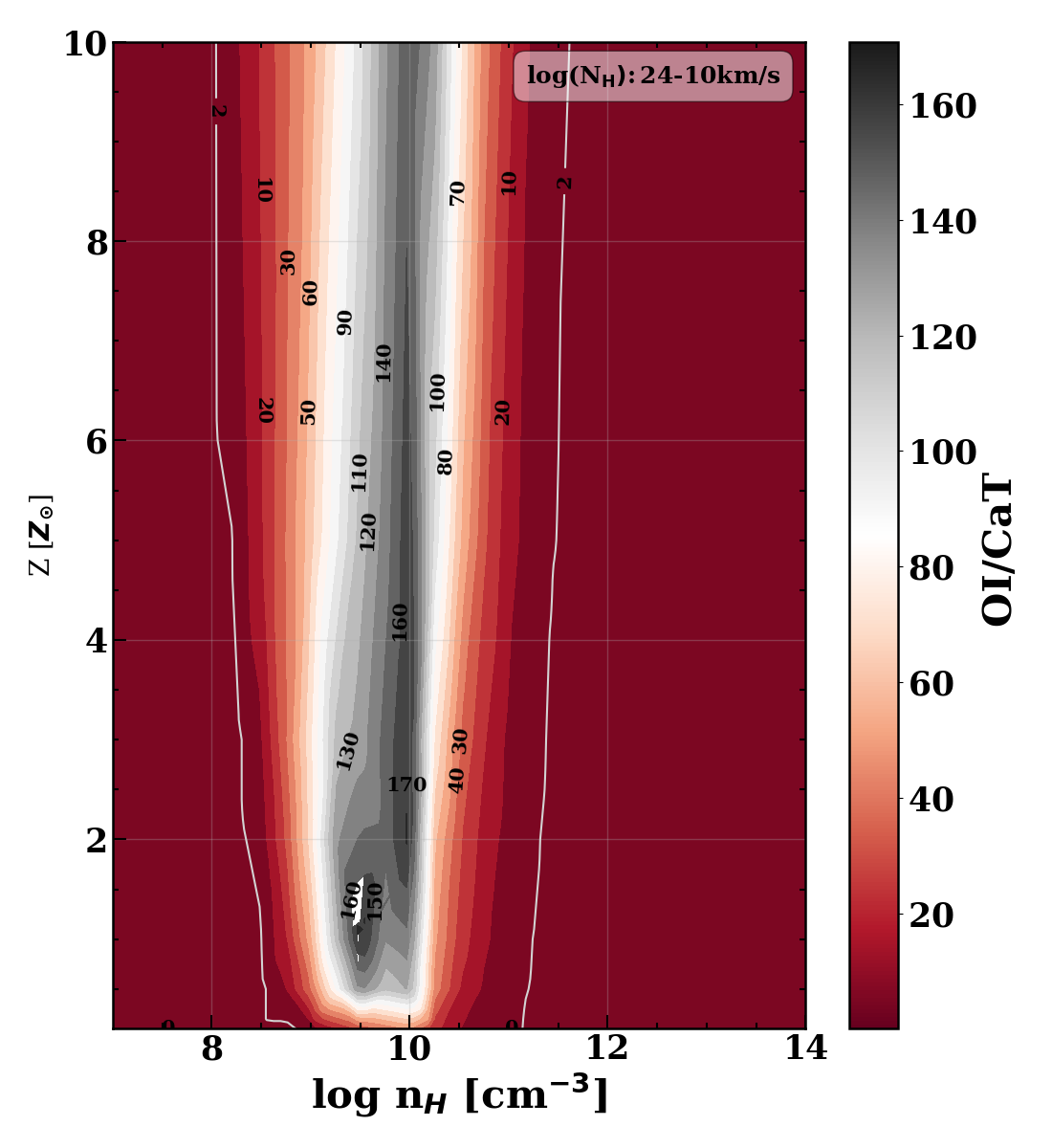}
  \caption{Parameter phase composed by local hydrogen density (log (n$_{H}$) on the x-axis and metallicity in solar units on the y-axis. The color bars represent the O\,{\sc i}/CaT ratios from our models, and the white solid line (value equal 2) is the observed ratio. We used the parameter set of 10\,\kms{} for micro-turbulence and $10^{24}$\,cm$^{-2}$ for the column density.}
  \label{fig:Rcat_ROI}
\end{figure}

Since O\,{\sc i}, and CaT ions have been suggested as alternatives for the \feii{} study \citep{Panda_2021_CaFe-II, Martinez-Aldama_etal_2021}, we tested the observed O\,{\sc i}/CaT ratio in our models to determine their agreement with the theoretical predictions. The CaT and O\,{\sc i} observational constraints used in this work are derived from the I~Zw~1 spectrum fitting published in \citet{dias2022properties}. Figure~\ref{fig:Rcat_ROI} shows the O\,{\sc i}/CaT ratio as a function of metallicity ($Z_{\odot}$) and local hydrogen density. The contour lines indicate different ratio values, with colors representing the intensity. Our observed ratio of O\,{\sc i}/CaT is represented by the white solid-line contour on the plot, corresponding to a range from sub-solar metallicity, and densities n$_{H}=10\,^{9.0}\,-\,10\,^{11.75}\,{\rm cm}^{-3}$. This shows that the physical conditions in these areas of the BLR are comparable to those where \feii{} is formed. The very good agreement between our observations and theoretical predictions supports the hypothesis that O\,{\sc i} and CaT emissions arise under similar physical conditions as \feii{}, reinforcing the idea that these ions can be used as proxies for studying \feii{} emissions in AGNs, as proposed by \citet{Panda2020OpticalModelling, Panda_2021_CaFe-II, Martinez-Aldama_etal_2021}.

Besides the overall agreement, we note that O,{\sc i} and CaT show differences in their physical ranges at their peak emission intensities. Figure~\ref{fig:apendicec} presents the luminosity of O\,{\sc i} (panel a), CaT (panel b), the sum of \feii{} emissions at 1\,$\mu$m (panel c), and luminosity of the 4570~\AA\ bump (panel d), as functions of local hydrogen density (n$_{H}$) and metallicity ($Z_{\odot}$). The observed line luminosity values are represented by the solid black line. Overall, the observed luminosities are well reproduced within the density range of 10$^{10.0}$ to 10$^{12.0}$ cm$^{-3}$. From Figure~\ref{fig:apendicec} we may see that the physical conditions necessary to reach the maximum luminosity of the CaT gas do not coincide with those necessary to reach the maximum NIR \feii{} and O\,{\sc i} luminosities.

In Fig.~\ref{fig:apendicec} panel (a), the maximum luminosity of O\,{\sc i} occurs within the density range of 10$^{10.0}$ to 10$^{11.0}$ cm$^{-3}$, with observed values within the sub-solar to solar metallicity range. This indicates that our models can accurately reproduce these observations at their maximum emitted luminosity. The density range aligns with the region where \feii{} NIR emission is also maximized, suggesting that these emissions likely occur under the same physical conditions or even in co-spatial regions within the BLR. 

While the observed CaT luminosity can be reproduced at densities similar to those of O\,{\sc i} and NIR \feii{} emissions (around 10$^{10.0}$ to 10$^{11.0}$ cm$^{-3}$), the maximum CaT luminosity is achieved at higher densities, starting from 10$^{11.0}$ cm$^{-3}$. Notably, the highest luminosities for optical \feii{} are also found at densities comparable to those where the peak CaT emission occurs (see Fig.~\ref{fig:apendicec}, panel (d)). These results suggest an alternative scenario in which the majority of CaT and \feii{} emission originates from the denser and more internal regions of the BLR, corroborating previous findings \citep{Panda2020OpticalModelling, Panda_2021_CaFe-II}. In this possible scenario, CaT and \feii{} emissions could be produced at similar densities to those of O\,{\sc i}, and NIR \feii{}, but their maximum emissions are expected to come from a denser and deeper region of the BLR.

\subsection{ Emission lines emissivities }

Previous sections have shown that with the present model that employs the Smyth \feii{} dataset, it is possible to reproduce the optical and NIR \feii{} intensities observed in I~Zw~1. With local hydrogen density conditions ranging from $10^{11.0}$ to $10^{12.0}\,\mathrm{cm}^{-3}$, metallicities up to solar, and the introduction of microturbulence ($v_{\text{turb}} = 10\,\mathrm{km\,s^{-1}}$), we significantly improve the agreement between the observed and simulated emission line intensity. Based on these results, we selected representative physical conditions $10^{9.5}$ $<$ n$_{\rm H} < $ $10^{13.5}\,\mathrm{cm^{-3}}$, and 0.5$\,Z_{\rm \odot} \leq Z \leq$ 5$\,Z_{\rm \odot}$, to explore the emissivity distribution throughout the cloud for each emission line. The other cases of emissivity profiles are hosted on Zenodo\footnote{\href{https://zenodo.org/records/16053381?preview=1&token=eyJhbGciOiJIUzUxMiJ9.eyJpZCI6ImMzOTc0OTZjLTYwNmMtNGM5Zi1hNjI3LWM3NGU5YTcyMGZlOSIsImRhdGEiOnt9LCJyYW5kb20iOiI0MzViY2ZkNzNlMjNlYWIwMjQwZjUwY2U2MzUzNjQ1NyJ9.nM7cwLD9A2XCJQsu0KFQ-fqyhXw7RlSMSfr5ayD5Zc1IgLksmXJuSsFQPe2BL4oiBu8as504wnCsr7-NACRadg}{https://zenodo.org/records/16053381}}.

\begin{figure*}
    \centering
    \includegraphics[width=\textwidth]{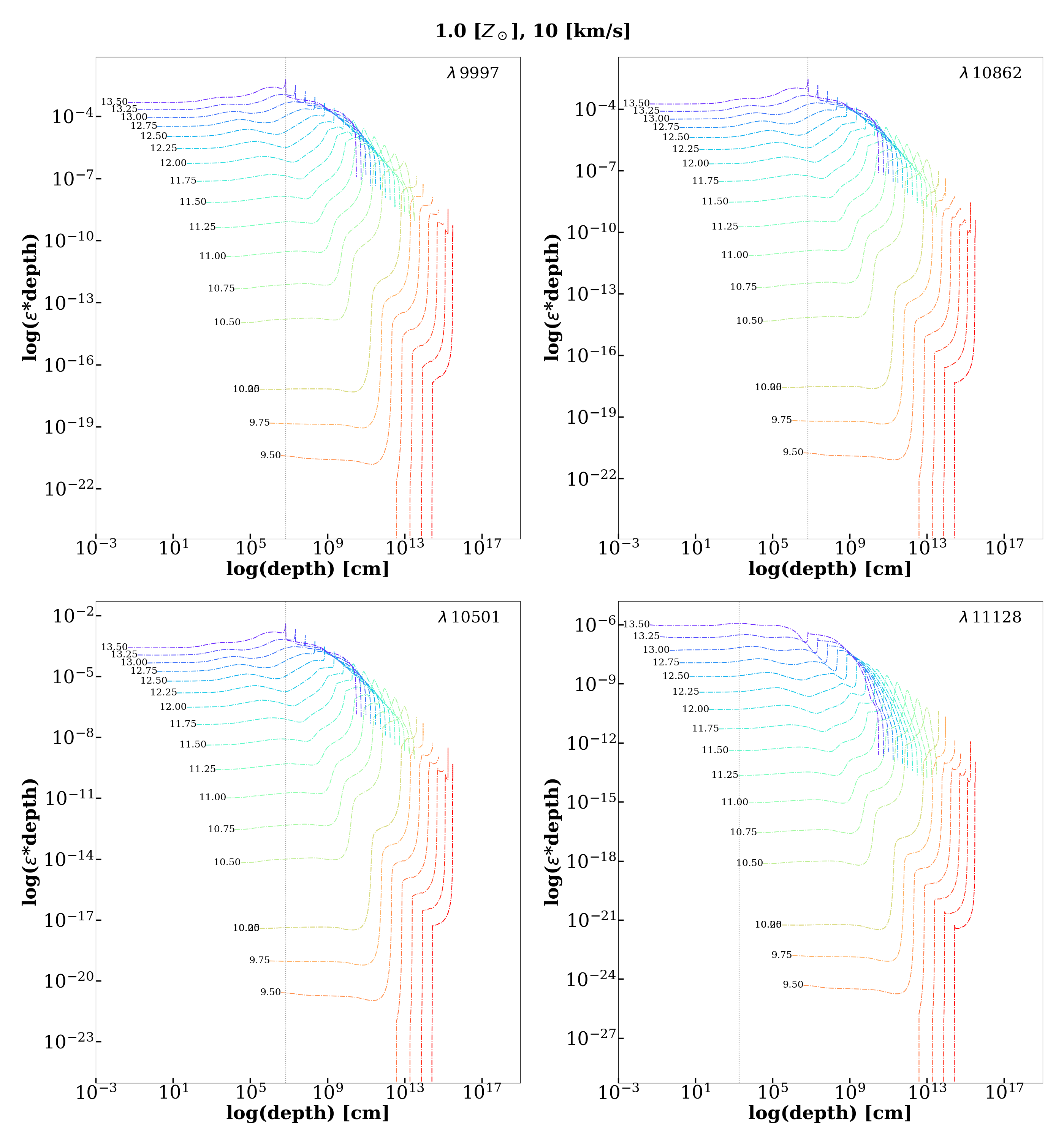}
    \caption{Emissivity of the four 1-micron \feii{} emission lines. Different colors mean different local hydrogen densities. All cases assume solar metal content (Z$_{\odot}$). The vertical lines indicate the location of the maximum value of emissivity multiplied by depth for the respective lines.}
    \label{fig:emissivity1}
\end{figure*}

Figure~\ref{fig:emissivity1} presents the emissivity profiles as a function of cloud depth for the four \feii{} NIR lines, assuming solar metallicity and the full hydrogen density range. In general, the results indicate that the $\lambda 9997$, $\lambda 10501$, and $\lambda 10862$ \feii{} lines originate from regions of similar depths within the cloud and reach their maximum emissivity at comparable depths. An exception is the $\lambda 11127$ line, which consistently reaches its peak emissivity at distinct depths from that of the other three lines, being strongly influenced by the metallicity (see Appendix~\ref{fig:emissivity2}). This behavior suggests that $\lambda 11127$ is particularly sensitive to the metal content of the gas and may not arise from the same regions as the other NIR \feii{} lines, as shown in Appendix~\ref{fig:emissivity2}. In contrast, the other NIR \feii{} lines consistently peak at similar depths over all metallicities, from sub-solar to five times solar metal content.

Furthermore, the optical \feii{} emission line at 4555\,\AA{} -- most prominent \feii{} line in the bump centered at 4570~\AA{} region -- shows a comparable behavior respect to the $\lambda 9997$, $\lambda 10501$, and $\lambda 10862$ \feii{} lines, with its maximum emissivity arising from closer parts (closer to the inner face of the cloud), reinforcing the idea that these transitions are formed in physically similar zones within the cloud.

To better understand the nature of the emitting regions and compare the behavior of different ions, we examined the depth within the cloud where the maximum emissivity occurs for each line as a function of local hydrogen density and metallicity (Figures~\ref{fig:emissivity1}, \ref{fig:emissivity2}, and \ref{fig:emissivity3}). This includes the \feii{} 1-micron lines, optical \feii{} 4555\,\AA, O\,\textsc{i} 8446\,\AA, and the Ca\,\textsc{ii} triplet (the NIR triplet emitting at $\lambda$8498\AA, $\lambda$8542\AA\, and $\lambda$8662\AA).

We note that the region of maximum emissivity for the 1-micron \feii{} lines lies mainly between $\log n_\mathrm{H} $ ($\mathrm{cm^{-3}}$) = 9.5 and 11.0, with emissivity increasing as the metallicity increases (see Appendix~\ref{fig:emissivity2}). For the $\lambda 11127$ line (Apendix~\ref{fig:emissivity2}), bottom-right panel), the ion exhibits a concentrated emissivity peak at $\log n_\mathrm{H} = 9.75$$\,\mathrm{cm^{-3}}$ for a metallicity of three times solar, whereas, $\lambda 9997$, $\lambda 10501$, and $\lambda 10862$ \feii{} lines reach their maximum emissivity across a broader range of physical conditions. These findings suggest that the $\lambda 11127$ line is likely maximized in different depths within the cloud compared to the other \feii{} NIR lines. Although all of these lines correspond to the same transition levels in the \feii{} ion, their emissivity profiles are not necessarily similarly affected by different physical conditions, further emphasizing the distinct excitation conditions of $\lambda 11127$.

Figure~\ref{fig:emissivity3} displays similar results for \feii{} 4555\,\AA, O\,\textsc{i} 8446\,\AA, and Ca\,\textsc{ii}. The optical \feii{} line peaks at higher metallicities ($Z \gtrsim 2Z_{\odot}$) and remains strong up to $\log n_\mathrm{H} \sim 11.5$ cm$^{-3}$, suggesting that its maximum emission can originate from a broader range of depths within the BLR cloud than its NIR counterparts. When comparing with O\,\textsc{i} and Ca\,\textsc{ii}, we find that the maximum of the O\,\textsc{i} 8446\,\AA{} emission comes from relatively closer emitting regions, in agreement with the \feii{} lines and sharing similar density and metallicity, reaching its peak at $Z \sim 3Z_{\odot}$. In contrast, the Ca\textsc{ii} emission behaves quite differently; its maximum emissivity region arises from deeper layers in the cloud under more extreme physical conditions.

It is worth noting, however, that despite its distinct peak location, Ca\,\textsc{ii} exhibits high emissivity values throughout the entire cloud and can surpass the emissivity of other lines at their respective maxima. This widespread and intense emission makes the comparison of Ca\,\textsc{ii} with \feii{} and O\,\textsc{i} intriguing. While Ca\,\textsc{ii} can be emitted efficiently under a broad range of conditions, its maximum appears in distinct physical regions, suggesting a different excitation environment with extreme physical conditions ($n_\mathrm{H} \gtrsim 10^{12}\,\mathrm{cm^{-3}}$).

\subsection{ The multifaceted perspective of \feii{} emission}

A complete understanding of the \feii{} emission across the broad range of wavelengths will enable us to characterize and classify various Type-1 AGNs on the Eigenvector 1 scheme -- which is closely related to the physical conditions of the line-emitting gas around the supermassive black hole \citep[see e.g.,][]{Marinello2016, Panda2018ModelingPlane, Marziani2018AQuasars,Panda2019TheOrientation,Floris_etal_2025}--and especially, is also crucial for tracing the chemical evolution of the AGN across cosmic time \citep[see e.g.,][]{hamannferland92, dietrich2003, shinetal13, Shin_etal_2021, Martinez-Aldama_etal_2021}.

As has been highlighted in \citet{Martinez-Aldama_etal_2021}, the $\alpha$-elements, e.g., Ca, Mg, O, etc., are predominantly produced by Type~II supernovae (SNe) after the explosion of massive stars (7 M$_\odot$ $< M_\star <100$ M$_\odot$) on timescales of $10^7$ yr, while  Fe is mostly produced by Type~Ia SNe from white dwarf progenitors on longer timescales $\sim1$ Gyr \citep{hamann-ferland1993}. In their study using Ca as a proxy for the $\alpha$-element and with a sample covering a redshift range [0.01, 1.68], \citet{Martinez-Aldama_etal_2021} reaffirm that the ratio Fe/Ca can be used as a clock for constraining the star formation, the metal content, and the age of the AGN \citep{mateucci1994, hamannferland92}.

These studies have primarily made use of the \feii{} emission observed in the optical and/or the UV regime, which, due to an immense number of transitions, makes up a pseudo-continuum of many blended multiplets. Due to the blending, an exact characterization of each \feii{} transition becomes a difficult task, and one mostly resorts to a composite analysis. These composite analyses have been extremely helpful to adjudge the connection of the \feii{} emission to the accretion state of the AGN, in addition to providing a rough estimate of the total iron content in the BLR \citep[see e.g.,][]{baldwin2004, Bruhweiler2008Modeling1, Panda2018ModelingPlane, Panda2019TheOrientation}.

Another far-reaching application of the optical \feii{} emission strength is its strong correlation to the accretion efficiency of the AGN. The quasar main sequence studies have revealed this connection \citep{Shen2014TheOverview, Marziani2018AQuasars, Panda_2024FrASS}, which has allowed us to incorporate the \rfeop{} as a direct observable from the observations to infer, not only the metal content and chemical state, but also the Eddington ratio of the said AGN \citep{Marziani2018AQuasars, Martinez-Aldama_etal_2021}. With the increasing number of AGNs that have been reverberation-mapped to infer the sizes of their BLR emitting regions encompassing a wide range in black hole masses and Eddington ratios \citep{Bentz2013ApJ, Grier2017TheCampaign, Du2018Mapping, Panda_Marziani_2023FrASS}, we have started to see a marked scatter in the classical BLR radius vs. AGN luminosity relation, i.e., R$_{\rm BLR}$ - L$_{5100\rm\AA}$ relation \citep{Kaspi2000ReverberationNuclei, Bentz2013ApJ}. Fortunately, \citet{Du2019TheNuclei} and \citet{Panda_Marziani_2023FrASS} have noted that the \rfeop{} parameter, when incorporated within the existing bi-variate relation between the BLR radius and AGN luminosity, can help alleviating the scatter in the relation. This helps us to utilize the refined R$_{\rm BLR}$ - L$_{5100\rm\AA}$ - \rfeop{} relation to infer luminosity distances independent of the cosmological model, and build the Hubble-Lema\^itre diagram with AGNs extending to high redshifts \citep{Watson2011ApJ, Haas2011, Czerny2021,Czerny2023,Panda_Marziani_2023FrASS}. However, the aforementioned complications with estimating the \rfeop{} and its UV counterpart, and the success in recovering the similarity in the \rfenir{}, have opened new avenues to exploit the correlation and extend the analysis of these \feii{}-rich AGNs in the NIR.




\section{Conclusions and Summary}\label{sec:5}

In this work, we conducted a comprehensive study of the \feii{} emission across the optical and near-infrared (NIR) regimes. Our analysis employs state-of-the-art photoionization modeling and atomic \feii{} datasets to explore the physical conditions of the \feii{} gas within the BLR in prototype NLS1 I~Zw~1, specifically targeting the reproducibility of the \rfeop{} and \rfenir{} \feii{} intensity. In addition, we test theoretical and observational predictions \citep{Marinello2016, Panda2020OpticalModelling} of compatibility with the \feii{} gas physical properties with the O\,{\sc i}, and CaT triplet (CaT) emitting gas. 

Our results confirm that both optical and NIR \feii{} emissions can be simultaneously reproduced under overlapping physical conditions when using the atomic datasets of \citet{smith19} and \citet{tayal2018}. For the former, successful modelling occurs at hydrogen densities between 10$^{11.0}$ and 10$^{12.0}$ cm$^{-3}$, with metallicities up to solar. The main distinction between the two datasets is that \citet{smith19} allows for a good match to both optical and NIR \feii{} features under slightly lower metallicities (close solar), whereas \citet{tayal2018} generally requires higher metallicities (up to 2~Z$_\odot$) to achieve similar results. 
We conclude that both updated atomic datasets — \citet{smith19} and \citet{tayal2018} — represent a significant improvement over the older \citet{Verner1999NUMERICALSPECTRA} dataset, as they successfully reproduce the observed \feii{} emissions under less extreme physical conditions. 

Moreover, the comparison between the Fe\,{\sc ii} measurements derived from the spectrum of \citet{diasdossantos24} and those obtained from the new I\,Zw\,1 spectrum reveal us that both observational constraints intersect in a common region of parameter space in the simulations, indicating that our main conclusions regarding the physical conditions of the emitting gas remain robust -- independently of the spectral epoch so far in the I\,Zw\,1 case. 

In addition, the introduction of microturbulence (10~\kms{}) in the models significantly improves the agreement between the observed and simulated spectra, producing effects similar to an increase in metallicity and bringing the theoretical predictions closer to the observed values \citep{Panda2020OpticalModelling}.

Furthermore, our work highlights the crucial role of Lyman-$\alpha$ fluorescence (Ly$\alpha$) in boosting \feii{} NIR emission, particularly the 1\,$\mu$m lines, confirming that this excitation mechanism contributes significantly to the \feii{} emission in the UV, optical, and NIR spectra. However, the influence of Ly$\alpha$ on optical emissions, such as the 4570~\AA\ bump, is less pronounced, suggesting that other mechanisms might be at play in these regions, such as collisional excitation.

We observed that the  O\,{\sc i}/CaT ratio agrees with theoretical models, suggesting that O\,{\sc i} and CaT emissions can serve as analogs for \feii{} emissions, particularly under similar physical conditions. However, we also found that while the observed CaT luminosity can be reproduced at densities comparable to those of O\,{\sc i} and NIR \feii{} emissions, the maximum CaT luminosity is observed at higher densities, similar to the maximum of optical \feii{}.

These findings may suggest that CaT and optical \feii{} emissions originate mainly from denser inner regions of the BLR, while O\,{\sc i} and NIR \feii{} emissions may arise mainly from relatively less dense outer regions. This possible scenario suggests a stratified structure in the BLR, in which different ions peak their maximum emission at different depths. Such a stratification within the BLR hints at a complex layering of emission regions, with O\,{\sc i}, CaT, and \feii{} emissions tracing different physical conditions within the same radial distance but possibly at varying vertical distance above the accretion disk.  This understanding supports the idea that, while O\,{\sc i} and CaT can act as proxies for \feii{} under certain conditions, the exact physical context must be considered to accurately interpret these emissions.

Our study provides a more detailed view of the physical conditions within the low ionization gas, emphasizing the importance of considering an appropriate atomic dataset for the spectral region of interest. Future work could further refine these models, exploring other \feii{} emitter AGNs to verify if these findings can be generalized for I~Zw 1-like sources.

\begin{acknowledgments}
The authors thank the Brazilian Agencies: Agency of Coordena\c{c}\~ao de Aperfei\c{c}oamento de Pessoal de N\'{\i}vel Superior (CAPES), and Conselho Nacional de Desenvolvimento Cient\'{\i}fico e Tecnol\'ogico (CNPq). This study was financed in part by the Coordenação de Aperfeiçoamento de Pessoal de Nível Superior $-$ Brasil (CAPES) $-$ Finance Code 001. We thank Marios Chatzikos and the \cloudy{} team for giving us access to the developmental version. We thank the anonymous referee for the constructive comments and suggestions. SP is supported by the international Gemini Observatory, a program of NSF NOIRLab, which is managed by the Association of Universities for Research in Astronomy (AURA) under a cooperative agreement with the U.S. National Science Foundation, on behalf of the Gemini partnership of Argentina, Brazil, Canada, Chile, the Republic of Korea, and the United States of America.
\end{acknowledgments}

%

\vspace{5mm}
\facilities{IRTF, CASLEO:JST}


\software{numpy \citep{numpy}, matplotlib \citep{matplotlib}, Cloudy \citep{Ferland1998CLOUDYSpectra, Ferland2017TheCloudy}}



\bibliography{reference}{}
\bibliographystyle{aasjournal}

\appendix
\restartappendixnumbering
\section{Supplementary Figures and Tables}

\begin{table}
\centering
\caption{Comparison between Fe\,{\sc ii} measurements in the $1\,\mu$m region from \citet{diasdossantos24} and the new I\,Zw\,1 spectrum. Fluxes are given in units of $10^{-14}$ erg s$^{-1}$ cm$^{-2}$.}
\label{tab:feii_comparison}
\begin{tabular}{lcccccc}
\hline
 & Fe\,{\sc ii} $\lambda$9998 & Fe\,{\sc ii} $\lambda$10502 & Fe\,{\sc ii} $\lambda$10863 & Fe\,{\sc ii} $\lambda$11127 &  Pa$\gamma$ (broad) & $R_{1\mu{\rm m}}$ \\
\hline
Previous work & $4.13 \pm 1.06$ & $2.70 \pm 1.38$ & $0.15 \pm 0.17$ & $2.32 \pm 1.29$ &  $5.96 \pm 0.83$ & $0.64$--$1.55$ \\
New spectrum  & $2.40 \pm 0.09$ & $2.34 \pm 0.09$ & $1.75 \pm 0.06$ & $1.17 \pm 0.04$ & $9.02 \pm 0.45 $ & $0.35$--$0.85$ \\
\hline
\end{tabular}
\end{table}

\begin{figure*}[h]
\centerline{\includegraphics[width=\textwidth]{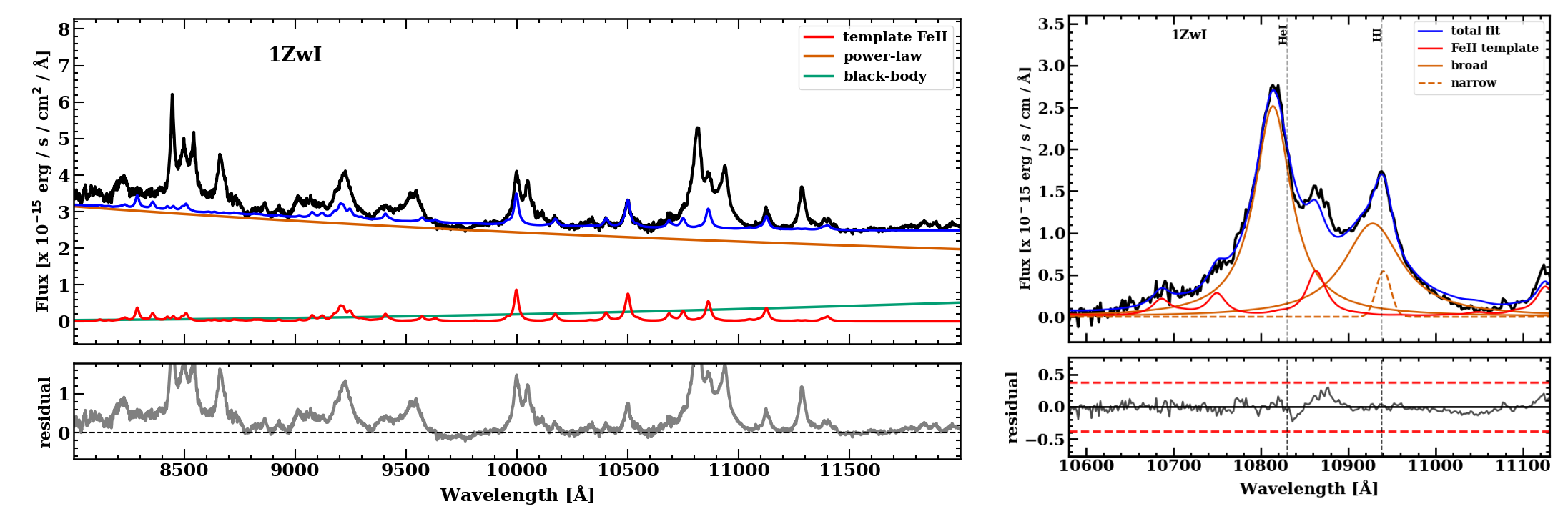}}
\caption{ Left panel:} Continuum decomposition and line fitting procedure carried out in I\,Zw\,1 near-infrared spectrum (in black). The continuum components are in green and orange, respectively, for the host dust and accretion disk contributions. The Fe\,{\sc ii} semi-empirical \cite{Garcia-Rissmann2012AGALAXIES} template is fitted in red, and the blue curve represents the sum of all components. {Right panel:} The Pa$\gamma$-He{\sc i} line blended with He{\sc i} for the observed spectrum (in black)  and the fits to the BLR (solid orange) and NLR components. See text for details. Dashed lines represent Gaussian components of the NLR, while solid lines represent BLR profiles (Gaussian or Lorentzian).

\end{figure*} \label{new_spec}

\begin{figure}[h]
\centering
{\includegraphics[width=\textwidth]{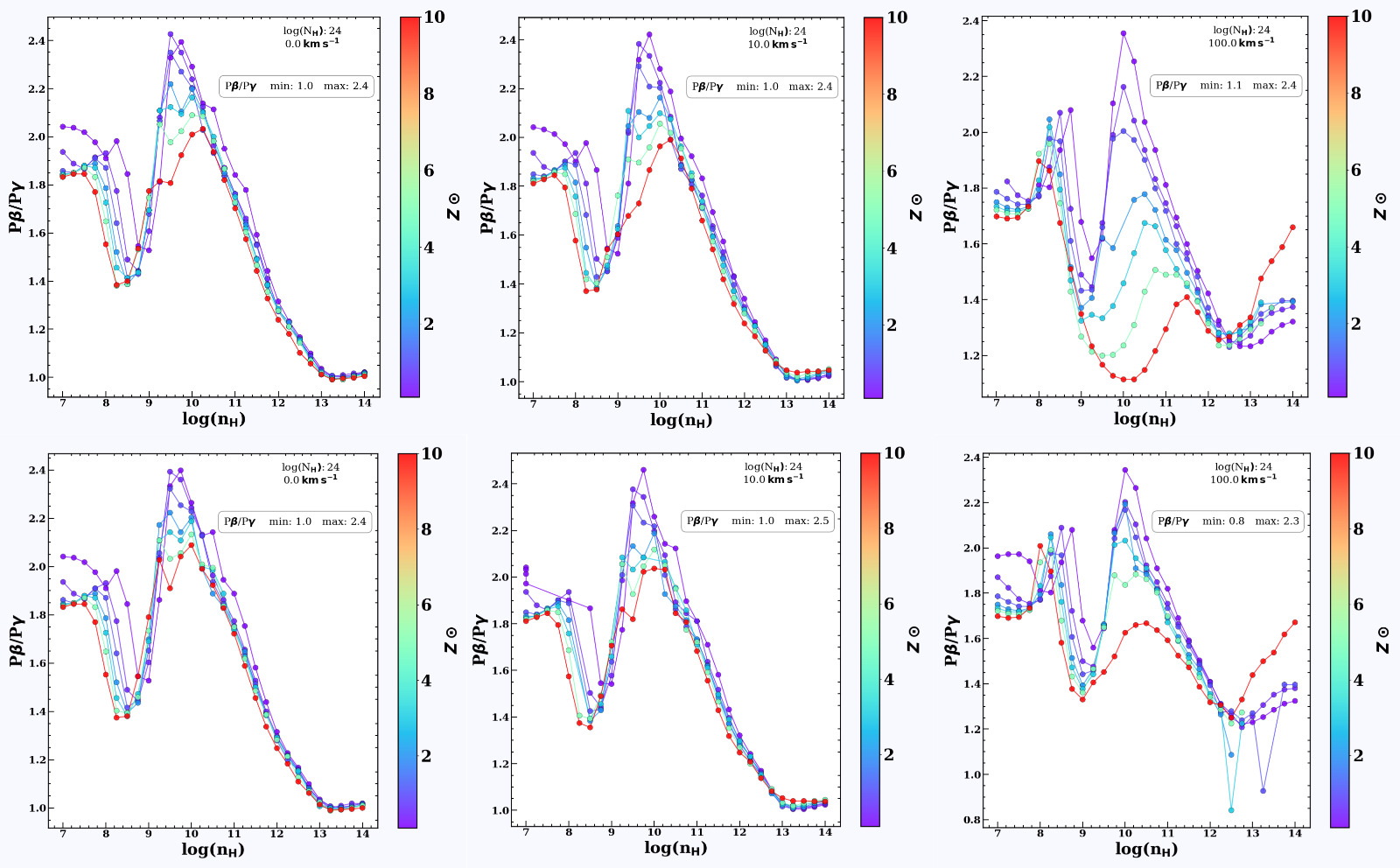}}

\caption{The luminosity ratio of Pa$\beta$ to Pa$\gamma$ obtained from the simulations as a function of the local hydrogen density, n$_{\rm H}$. The color scale represents different metallicities. A fixed cloud column density of 10$^{24}$ cm$^{-2}$ is assumed. Panels (a) to (c) correspond to the \citet{smith19} atomic dataset, while panels (d) to (f) are based on \citet{tayal2018}. From left to right, the panels show the cases of zero, 10, and 100\,\kms{} microturbulent velocities, respectively.}

\label{apendicea}
\end{figure}

\begin{figure}[h]
\centering
{\includegraphics[width=\textwidth]{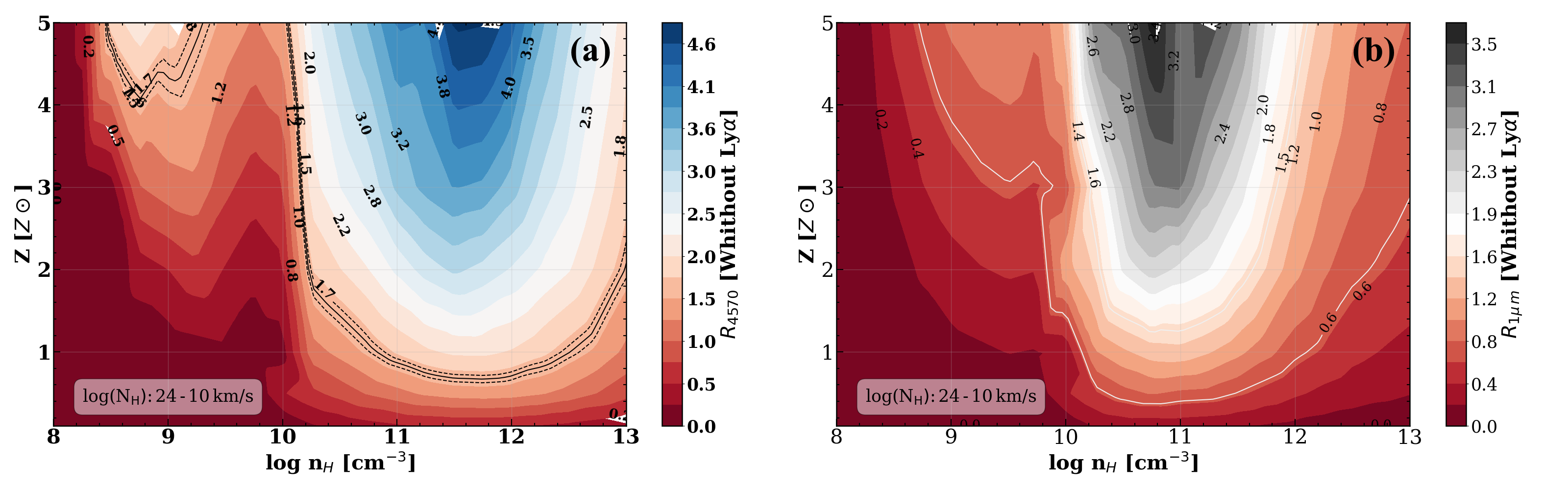}}
\caption{The figure shows the optical ratio $R_{4570}$ (top panel), and the $R_{1\mu m}$ (bottom panel). Both panels display the ratios for 10$^{8.0}$\,cm$^{-3}$$<\,$n$_{\rm H}\,<$10$^{13.0}$\,cm$^{-3}$. The colors correspond to different metal contents. The cloud column density is 10$^{24}$\,cm$^{-2}$ and the turbulence velocity is 10\,\kms{}.}
\label{apendiceb}
\end{figure}

\begin{figure*}
\centering
  \includegraphics[width=\textwidth]{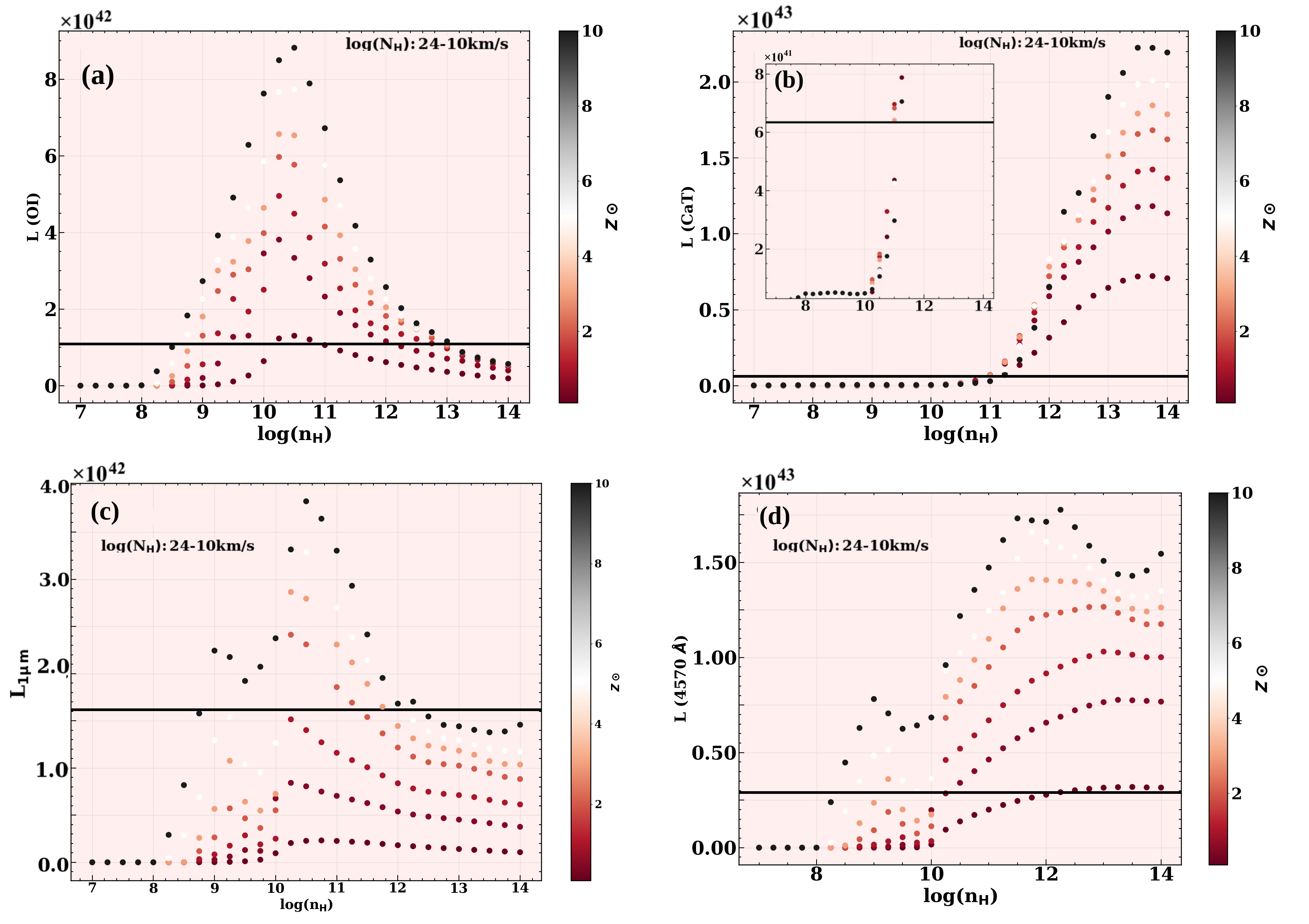}
  \caption{Parameter phase, composed by local hydrogen density (log (n$_{H}$) on the x-axis and luminosity on the y-axis. The color bars represent the metallicity in solar units, and the black solid line is the luminosity observed value. From panel (a) to (d), show O\,{\sc i} luminosity, Ca\,{\sc ii} luminosity, 1\,$\mu$m luminosity, and optical luminosity of 4570~\AA\ bump. For all panels, we use 10\,\kms{} micro-turbulence and $10^{24}$\,cm$^{-2}$ column density.}
  \label{fig:apendicec}
\end{figure*}








\begin{figure*}
    \centering
    \includegraphics[width=\textwidth]{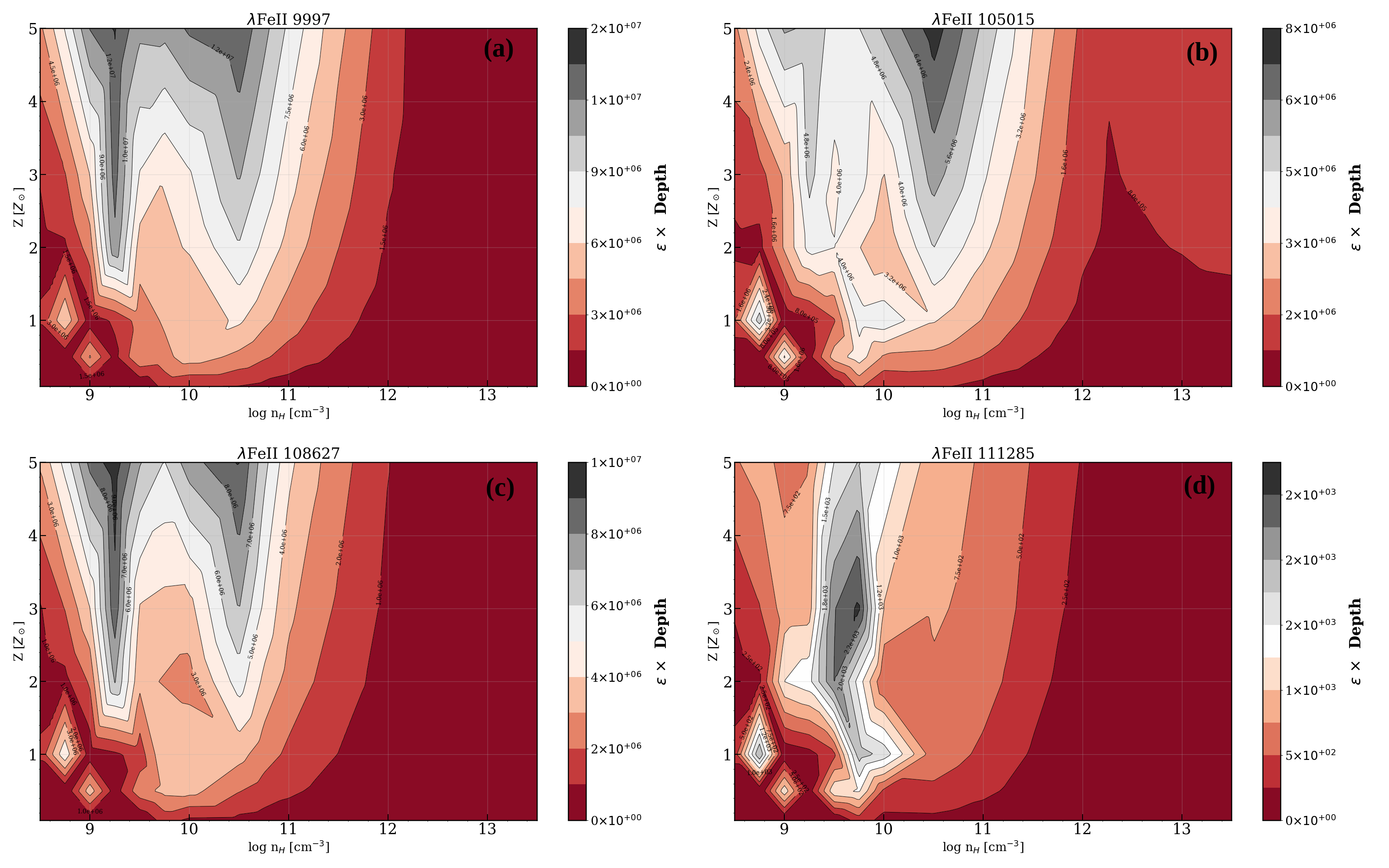}
    \caption{Emissivity of the four 1-micron \feii{} emission lines. Different colors mean the local hydrogen densities. }
    \label{fig:emissivity2}
\end{figure*}

\begin{figure*}
    \centering
    \includegraphics[width=0.85\textwidth]{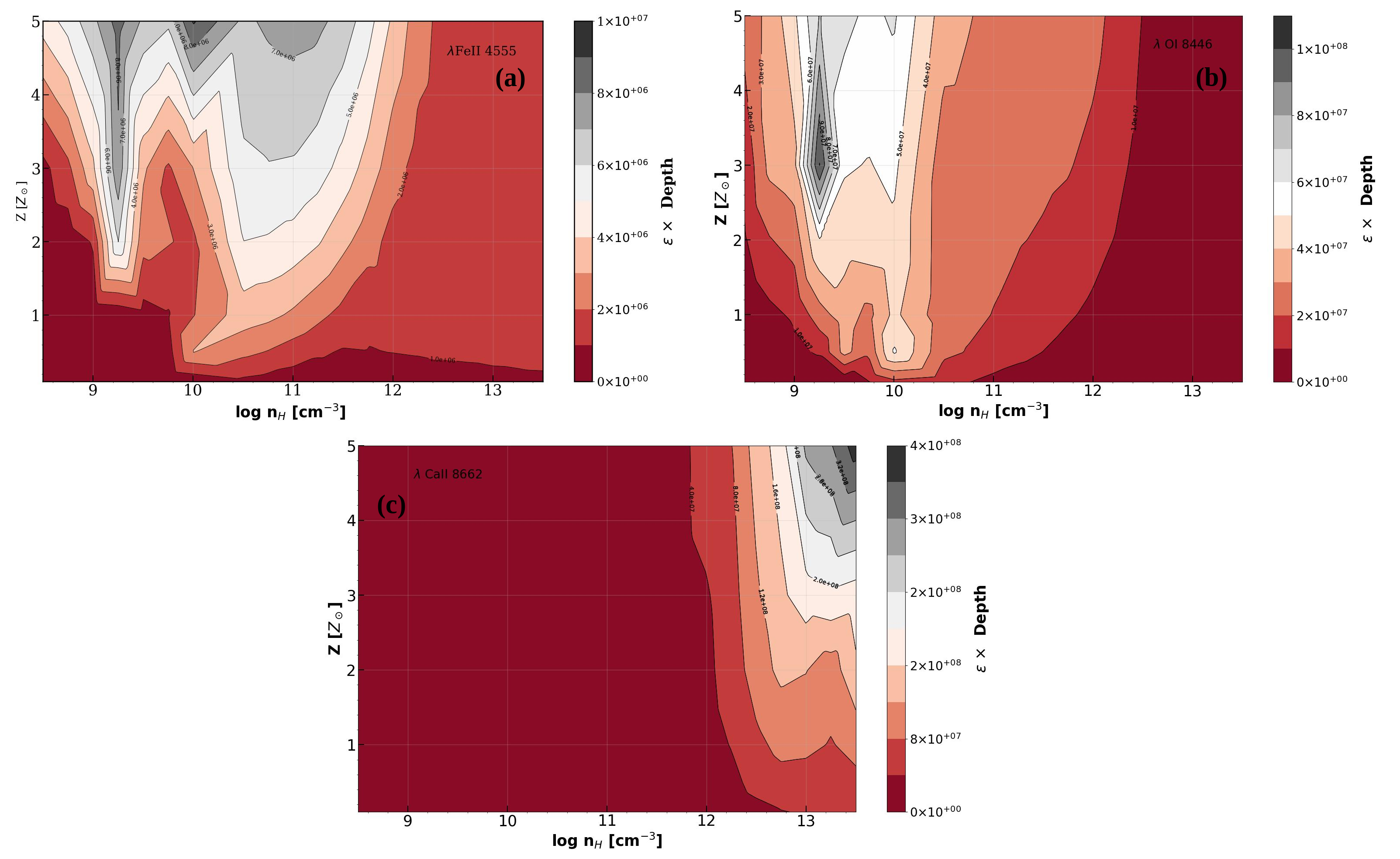}
    \caption{Similar to Figure~\ref{fig:emissivity2}, but for optical \feii{}, O\,{\sc i} and CaT lines. }
    \label{fig:emissivity3}
\end{figure*}

\end{document}